\documentclass[11pt,aps,nofootinbib,prd,aps,epsf,floats,axodraw,amsmath,amssymb,amsfonts]{revtex4}
\usepackage{amsmath, amssymb}
\newcommand{\mathsym}[1]{{}}

\usepackage{graphicx}
\usepackage{amsmath}
\usepackage{amssymb}
\usepackage{bm}
\newcommand{\be}{\begin{equation}}
\newcommand{\ee}{\end{equation}}
\newcommand{\bea}{\begin{eqnarray}}
\newcommand{\eea}{\end{eqnarray}}

\newcommand{\rem}[1]{}
\newsavebox{\PSLASH}
 \sbox{\PSLASH}{$p$\hspace{-1.8mm}/}
 
\renewcommand{\theequation}{\thesection.\arabic{equation}}
\newcounter{saveeqn}
\newcommand{\add}{\addtocounter{equation}{1}}
\newcommand{\alpheqn}{\setcounter{saveeqn}{\value{equation}}%
\setcounter{equation}{0}%
\renewcommand{\theequation}{\mbox{\thesection.\arabic{saveeqn}{\alph{equation}}}}}
\newcommand{\reseteqn}{\setcounter{equation}{\value{saveeqn}}%
\renewcommand{\theequation}{\thesection.\arabic{equation}}}

 \newsavebox{\notrightarrow}
 \sbox{\notrightarrow}{$\to$\hspace{-4mm}/}
 
 \newsavebox{\PARTIALSLASH}
 \sbox{\PARTIALSLASH}{$\partial$\hspace{-1.6mm}/}
 
 \newsavebox{\ASLASH}
 \sbox{\ASLASH}{$A$\hspace{-2.1mm}/}
 
 \newsavebox{\KSLASH}
 \sbox{\KSLASH}{$k$\hspace{-1.8mm}/}
 
 \newsavebox{\LSLASH}
 \sbox{\LSLASH}{$\ell$\hspace{-1.8mm}/}
 
 \newsavebox{\QSLASH}
 \sbox{\QSLASH}{$q$\hspace{-1.8mm}/}
 
 \newsavebox{\DSLASH}
 \sbox{\DSLASH}{$D$\hspace{-2.2mm}/}
 
 \newsavebox{\DbfSLASH}
 \sbox{\DbfSLASH}{${\mathbf D}$\hspace{-2.8mm}/}
 
 \newsavebox{\DELVECRIGHT}
 \sbox{\DELVECRIGHT}{$\stackrel{\rightarrow}{\partial}$}
 
 \newcommand{\blue}{\IfColor{\textCadetBlue}{}}
\newcommand{\black}{\IfColor{\textBlack}{}}
\newcommand{\red}{\IfColor{\textRed}{}}
\newcommand{\green}{\IfColor{\textOliveGreen}{}}
\newcommand{\lila}{\IfColor{\textRedViolet}{}}

\begin{document}
\begin{flushright}
 [math-ph]
\end{flushright}
\title{Covariant Star Product on Semi-Conformally Flat Noncommutative\\ Calabi-Yau Manifolds and Noncommutative Topological Index Theorem}

\author{Amir Abbass Varshovi}\email{ab.varshovi@sci.ui.ac.ir/amirabbassv@ipm.ir/varshoviamirabbass@gmail.com}

\affiliation{Faculty of Mathematics and Statistics, Department of Applied Mathematics
and Computer Science, University of Isfahan, Isfahan, IRAN.\\
School of Mathematics, Institute for Research in Fundamental
Sciences (IPM), P.O. Box: 19395-5746, Tehran, IRAN.}
\begin{abstract}
\textbf{Abstract\textbf{:}} A differential geometric statement of the noncommutative topological index theorem is worked out for covariant star products on noncommutative vector bundles. For start, a noncommutative manifold is considered as a product space $X=Y \times Z$, wherein $Y$ is a closed manifold, and $Z$ is a flat Calabi-Yau $m$-fold. Also a semi-conformally flat metric is considered for $X$ which leads to a dynamical noncommutative spacetime from the viewpoint of noncommutative gravity. Based on the Kahler form of $Z$ the noncommutative star product is defined covariantly on vector bundles over $X$. This covariant star product leads to the celebrated Groenewold-Moyal product for trivial vector bundles and their flat connections, such as $C^\infty(X)$. Hereby, the noncommutative characteristic classes are defined properly and the noncommutative Chern-Weil theory is established by considering the covariant star product and the superconnection formalism. Finally, the index of the $\star$-noncommutative version of elliptic operators is studied and the noncommutative topological index theorem is stated accordingly. \\
\noindent \textbf{Keywords\textbf{:}} Covariant Star Product, Semi-Conformally Flat Metric, Kahler Form, Calabi-Yau manifold, Superconnection, Chern-Weil Theory, Noncommutative Topological Index Theorem.
\end{abstract}

\pacs{math-phys} \maketitle


\section{Introduction}\label{introduction}

\par Intimate correlations of Einstein-Yang-Mills theories and topological structures of their underlying manifolds have been understood thoroughly over the last four decades.\footnote{See for example \cite{donaldson, don, witten, witten'}.} Physical effects such as instantons, consistent and covariant anomalies, holonomy structures, Schwinger terms, Chern-Simons theory, etc. in Yang-Mills theories, together with topological features of gravity and string theory including BTZ topological black holes, topological strings, mirror symmetry, topological dyons in anti-de Sitter Einstein-Yang-Mills theories, Witten-Vafa twist, magnetic monopoles and the Seiberg-Witten equations, etc.\footnote{See for example \cite{nakahara, tillmann2, labastida, nash, schwarz, shuryak, ring, katz, hollands, shnir} and the references therein.} are established to contain significant information about the geometry and the topology of spacetime. However, the situation is somehow unclear when the spacetime $X$ is a noncommutative space. This noncommutativity is mostly inspired by superstring developments, in which the $B$ background field induces a Groenewold-Moyal star product for the spacetime.

\par Despite of the elegant achievements of noncommutative geometry via the spectacular machineries of cyclic cohomology, spectral triples, Connes-Chern characters and the fascinating descriptions of the index theorem for noncommutative manifolds via a beautiful local index formula \cite{connes}, it seems there is a serious lack of topological interpretation for Groenewold-Moyal noncommutative Yang-Mills theories. In principle, although some of the physical effects of noncommutative Yang-Mills theories have been formulated within noncommutative framework, but their topological meanings have not been understood thoroughly yet.

\par Actually, this failure is due to the ambiguity of the de Rham cohomological counterparts of such algebro-geometric formulations in noncommutative geometry. But, actually, for the Groenewold-Moyal noncommutative Yang-Mills theories, wherein we essentially deal with substantially ordinary (commutative) manifolds, the noncommutative (cyclic) and the commutative (de Rham) settings are expected to be connected via a clear mechanism including the topological features. Such a formulation has not been clearly developed yet via a definit topological method. Actually, the deRham and cyclic cohomologies for commutative algebra are dual objects, but for noncommutative algebras there is not a clear definition for the de Rham cohomology as one may expect for noncommutative Yang-Mills theories.

\par The Seiberg-Witten map is in fact an attempt to overcome this problem via a string theoretic approach given for the Dirac-Born-Infeld actions of open strings \cite{seiberg-witten}.\footnote{See \cite{varshovi2, varshovi3} for some cohomological approach to figure out the problem.} Meanwhile, it has already been established \cite {cattaneo} that this map is also in intimate correlation with the celebrated Kontsevich's quantization formula for Poisson manifolds \cite{kontsevich}, hence possesses some significant information of the background geometry/topology. Although in \cite{seiberg-witten} the authors have demonstrated that the emergence of the Groenewold-Moyal or the ordinary product is related to the chosen Pauli-Villars or the point splitting regularization methods, the topological information of the noncommutative curvature and its Yang-Mills theory has not been disclosed entirely. On the other hand, some exact formulations of the Seiberg-Witten map show that when one transfers from noncommutative to the ordinary Yang-Mills fields, the spacetime geometry admits some smooth deformation which is manifestly figured out via a non-flat metric or an emergent gravity \cite {yang}.\footnote{See also \cite{banerjee2, banerjee1}.}

\par Thus, the topological features of the underlying geometry of noncommutative Yang-Mills theories are still obscure and need to be discovered or interpreted by some well understood framework such as the de Rham cohomology. On the other hand, the formulation of the Seiberg-Witten map is only well-defined over flat vector bundles. In fact, whenever the Yang-Mills geometry is nontrivial the Groenewold-Moyal product loses its consistency due to the appearance of transition functions or gauge transformations. In such situations the noncommutative curvature and its corresponding Yang-Mills action would not be well-defined. Consequently, it seems that although the Seiberg-Witten map could be considered as an appropriate machinery for formulating noncommutative Yang-Mills on flat manifolds (e.g. $\mathbb R^n$), but it must be modified covariantly for nontrivial geometries. Indeed, the emergence of gravity via a non-flat metric over a deformed underlying manifold reveals some clear evidences for this theoretical refinement.

\par Actually, as we will show below, it can be seen that the noncommutativity of quantum fields has to be defined upon the Yang-Mills geometry over the spacetime manifold. Thus, the noncommutativity would be deformed from the Groenewold-Moyal product for the vacuum field to some fairly rigorous covariant star product in presence of the matter fields.\footnote{See \cite{vitale et al.} for some related evidences.} By this modified quantization method we would have a well-defined quantum theory which carries all possible topological information as the ordinary Yang-Mills theory do. Therefore, it seems that a covariant redefinition of the Seiberg-Witten map is appropriate and mandatory from both physical and geometric viewpoints.

\par More generally, the correlation between topological aspects of a noncommutative manifold and the de Rham cohomology of the underlying (commutative) geometry is still unclear. As the most significant feature, the noncommutative index theorem has not have a bright interpretation via the differential geometric structures on the underlying (commutative) manifold yet. Hence, the noncommutative versions of characteristic classes given by replacing the ordinary (wedge) product of differential forms with noncommutative star product suffer from our ignorance of their topological interpretations, if there exist any. However, as we will show in the following, it seems that the covariant star product which leads to covariant noncommutative curvature and well-defined characteristic classes gives rise to a beautiful solution for the problem.

\par In principle, employing the covariant star product (which is roughly defined by replacing the partial derivative by the covariant derivative in the Groenewold-Moyal star product) a well-defined framework for noncommutative differential geometry is constructed which inherits the most important possible geometric/topological aspects of the de Rham cohomology. In principle, the essential motivation to extend the definition of Groenewold-Moyal product to the covariant star product for the space of smooth sections of an arbitrary vector bundle, is to find a concrete formulation of noncommutative gravity which admits the most possible symmetries of spacetime. It seems that such a formulation must involve a dynamical noncommutative spacetime due to a non-constant noncommutativity matrix which carry some intrinsic structure of spacetime. Such formulations has recently been provided via a beautiful approach of matrix model based on the string features of IKKT and IIB models \cite{steinacker}.

\par Following the main idea of \cite{steinacker}, the initial objective of this research is to define a covariant star product by rescaling the noncommutativity matrix with the connection form (and consequently with the curvature) at each point of spacetime. In fact, the main mathematical idea of covariant star product in this perusal stems from \cite{bayen1, bayen2}, which put an associative covariant star product on the tangent bundle of a flat Poisson manifold. However, the procedure fails whenever the vector bundle is nontrivial, hence non-flat. Hereon, we employ the flatness of a Calabi-Yau manifold $Z$ together with a semi-conformally flat structure on $X$ to generalize the formulation of \cite{bayen1} to any given vector bundle over $Z$.\footnote{There are other definitions of covariant star products. For example see \cite{chaichian-zet, chaichian-zet2, vassilevich, zumino, blohman, masmoudi, manolakos, sperling, heckman, steinacker2} and the references therein.} The algebraic structure is effectively associative, gauge invariant and diffeomorphism invariant on the (co)tangent bundle. Therefore, it will provide a consistent setting for noncommutative Yang-Mills and noncommutative gravity.\footnote{There are other methods for replacing the non-dynamical noncommutative spacetime with matrix model or fuzzy spaces of extra dimensions. See for example the introduction of \cite {manolakos0} for a full list of and extensive description for alternative noncommutative models for gravity.}

\par In fact, as we will show below both the noncommutative Yang-Mills and the noncommutative Einstein-Hilbert actions given by this covariant star product remain invariant after any gauge and general coordinate transformation. Thus, the covariant star product carry the most symmetries of the ordinary Yang-Mills and gravity. Hence, it is expected that the covariant star product would keep the topological information of the underlying geometries. Within this consistent setting, as we will show in the following discussions, it could be firmly established that the ordinary characteristic classes are surprisingly cohomologous to their noncommutative versions given with the covariant star product. Afterward, a noncommutative Chern-Weil theorem will be obtained which would provide a spectacular topological achievement for noncommutative geometry. This will lead naturally to reliable evidence for the existence of some noncommutative topological index theorem. 

\par In this article we try to work out the entire expression of the noncommutative topological index theorem on noncommutative manifolds. Our strategy is first to generalize the Groenewold-Moyal star product for any given vector bundle over a noncommutative flat Calabi-Yau $m$-fold. We accomplish the process by replacing the partial derivative with the covariant derivative, coupled with the Kahler form of the base manifold, and furnishing the base manifold with a semi-conformally flat noncommutative structure. Next, by employing the superconnection method we will prove a noncommutative version of the celebrated Chern-Weil theorem which demonstrates an intimate correlation between topological properties of ordinary and noncommutative manifolds. Finally, the index of the noncommutative version of elliptic operators is studied and the full expression of the noncommutative topological index theorem is worked out on semi-conformally flat noncommutative Calabi-Yau manifolds. 


\section{Covariant Star Product on Associative Vector Bundles Over A Semi-Conformally Flat Noncommutative Calabi-Yau Manifold}
\setcounter{equation}{0}

\par Let $X$ be a $2n$-dimensional closed manifold which is described as a product space as $X=Y\times Z$, wherein $Y$, the so called \emph{commutative} component is itself an even dimensional closed orientable manifold, and $Z$, the \emph{noncommutative} part, is a flat Calabi-Yau $m$-fold equipped with a flat Kahler metric, say $h$, an almost complex structure $J$ and the Kahler form $\hat \omega$. In principle, to work out a noncommutative version of quantum gravity upon the viewpoints of the superstring theory or the supergravity, $X$ could be assumed as the spacetime with extra dimensions in $Z$. Here, $Z$ becomes the substructure of $X$ which is in fact, unobservable and responsible for the noncommutativity of quantum fields. Hence, $Y$ would be the observable four-dimensional spacetime. However, the fluctuations of the metric via the quantum gravity effects are not only considered for the commutative part $Y$.  

\par Actually, the definition of a Calabi-Yau manifold varies according to the documents, thus it must be specified that in the following we identify it as a Ricci flat Kahler manifold \cite{hubsh, gross}. Here, we generally put no condition on the canonical line bundle $\mathcal O_Z$ (e.g. its positive powers to be trivial) or on the holonomy group $\text{Hol}(h)$ (e.g. to be $SU(m)$). However, since $Z$ is a flat manifold it is seen that $\text{Hol}(h)$ is a discrete group, i.e. the group of deck transformations of its simply connected covering space \cite{hatcher, poor}.

\par Hence, in our investigation, a flat Calaby-Yau manifold is in fact a flat Kahler manifold. In fact, the collection of such spaces includes, of course, even dimensional tori and some special kinds of Kummer surfaces $K_3$ \cite{kronheimer, ohsawa}. But, in principle, flat Kahler manifolds could also be of more varied topologies with different algebro-geometric properties such as the Hodge numbers.\footnote{See \cite{flat-kahler} for more discussion about different types of flat Kahler manifolds of 4 and 6 dimensions.} In general, $Z=\frak G \setminus E(2m)/O(2m)$, where $E(k)=O(k) \ltimes \mathbb{R}^k$ is the Euclidean group of $k$-dimensional motions and $\frak G$ is a compact, discrete and torsion free subgroup of $E(2m)$. In this case, upon the Bieberbach's theorems we see that $\pi_1(Z)=\frak G$ and $\text{Hol}(h)=\frak G/T$, for $T\subset \frak G$ the subgroup of all pure translations \cite{bieberbach1, bieberbach2, flat-kahler}.\footnote{See \cite{ ratcliffe} for more discussion about the Bieberbach's theorems.}

\par To specify the spin structure of $X$ it is enough to assume that $Y$ is a spin manifold, since in fact, any Calaby-Yau manifold $Z$ is itself a spin manifold due to the equality of: $w_2(Z)=c_1(Z)~(\text{mod}~2)$, for $w_2(Z)$ the second Stiefel-Whitney class of $TZ$, and for $c_1(Z)=0$, the first Chern class of $TZ$ \cite{spin-geom}. Therefore, since $w_2(Y\times Z)=w_2(Y)+w_2(Z) +w_1(Y) \wedge w_1(Z)~(\text{mod}~2)$, then $X$ would be a spin manifold, whenever the first Stiefel-Whitney class of $TY$ (resp. $TZ$) vanishes identically, i.e. $w_1(Y)=0$ (resp. $w_1(Z)=0$).

\par We should continue with a statement of our main conventions about the noncommutative manifold $X$. Set $\dim Y=D$ and let $\mathcal A ^X=\{ (U^X_\alpha=U^Y_\alpha \times U^Z_\alpha \subset Y \times Z, \{x^\mu_\alpha  \}_{\mu=0}^{2n-1}  ) \}_{\alpha \in \frak I}$ be a covering atlas on $X$, whose the coordinate functions $x^\mu_\alpha$s split $X$; that is $x^a_\alpha$ belongs to $Y$ for $a=0, \cdots, D-1$, and $x^i_\alpha$s are local coordinates on $Z$ for $D\leq \mu \leq 2n-1$. As a convention, the local coordinates of $Y$ (resp. $Z$) are labeled with letters $a, b, c, \cdots$ (resp. $i, j, k, \cdots$) and referred to as the \emph{commutative coordinates} (resp. \emph{noncommutative coordinates}). However, the local coordinate functions of $X$ are generally labeled with Greek letters $0\leq \mu, \nu, \sigma, \cdots < 2n$. Thus, $(U^Y_\alpha,\{x^a_\alpha \})$ (resp. $(U^Z_\alpha,\{x^i_\alpha \})$) is a local coordinate system on $Y$ (resp. $Z$). But, however, if there is no confusion we drop the subscript $\alpha$ and simply write $x^\mu$, otherwise we keep $\alpha$. 

\par We assume that each $(U^Z_\alpha, \{ x^i_\alpha\})$ is a Darboux's chart for the symplectic form $\hat \omega$. This, ensures that the Kahler form $\hat \omega = \theta^\alpha_{ij} \text{d}x^i_\alpha \wedge \text{d}x_\alpha^j$ has constant coefficients $\theta_{ij}^\alpha$ on each $(U_\alpha, \{ x^i_\alpha \})$. Indeed, $\hat \omega =\sum_{i=1}^m\text{d}x^i_\alpha \wedge \text{d}x^{i+m}_\alpha$. In general, when there is no confusion we may drop the superscript $\alpha$ and simply write $\theta_{ij}$. However, we can always choose $x^i_\alpha$s so that $\theta_{ij}^\alpha = \theta^\beta_{ij}\in \mathbb R$, for each $\alpha, \beta \in \frak I$ with $U^Z_\alpha \bigcap U^Z_\beta \ne \emptyset$. Actually, according to the Bieberbach's theorems \cite{bieberbach1, bieberbach2} the flatness of $Z$ ensures that it is finitely covered by $\mathbb T^{2m}$. More precisely, since $Z$ is flat it is a quotient of $\mathbb T^{2m}$ modulo some finte group, i.e. the deck transformation group. Therefore, $Z$ is itself a Lie group, and hence possesses a trivial tangent (resp. cotangent) bundle; $TZ=Z \times \mathbb R^{2m}$ (resp. $T^*Z=Z \times \mathbb R^{2m}$). Thus, since $Z$ is Abelian it admits $2m$ numbers of globally defined nonvanishing vector fields $\{\partial_i\}_{i=1}^{2m}$ (resp. one-forms $\{ \text{d}x^i \}_{i=1}^{2m})$ with constant symplectic form $\hat \omega$ on the second component $\mathbb R^{2m}$. In fact, these $2m$ vector fields (resp. one-forms) are integrable locally on each contractible $U^Z_\alpha$, producing local coordinate functions $\{x^i_\alpha\}_{i=1}^{2m}$. Then, $\frac{\partial}{\partial x^i_\alpha}=\frac{\partial}{\partial x^i_\beta}=\partial_i$ (resp. $\text{d}x^i_\alpha =\text{d}x^i_\beta=\text{d}x^i$) on $U^Z_\alpha \bigcap U^Z_\beta \ne \emptyset$, and we obtain the globally defined $\hat \omega=\theta_{ij} \text{d}x^i \wedge \text{d}x^j$ with constant coefficients $\theta_{ij}$.\footnote{Conversely, one can similarly figure out the topology of $X$ as a principal $Z$-bundle over $Y$. In fact, $\{ \partial_i\}$ generates a smooth involutive distribution $\mathcal D$ on $X$ and its maximal connected integral manifolds provide the fibers of $X$ according to the Frobenius theorem \cite{warner}. The triviality of the topology of $X$, as a fibration over the base manifold $Y$, is indeed an immediate consequence of definition of $\mathcal D$ which admits a global basis. This fixes the unique differential and topologic structure of $X$ \cite{taubes}.}  

\par The open sets $U^Y_\alpha$ and $U^Z_\alpha$ (hence $U^X_\alpha$) are assumed to be contractible. The flat Kahler metric $h$, as introduced above, is the unique flat $J$ compatible Riemannian metric on $Z$ due to the Yau's theorem \cite{yau}. The points of $X$ are simply shown with $(y,z)\in Y\times Z$. The metric of $Y$, shown with $g_{ab}(y,z)$, is considered to be \emph{semi-conformally flat via noncommutative part $Z$}, i.e. $g_{ab}(y,z)=e^{\sigma(z)} G_{ab}(y)$, where $\sigma(z)$ belongs to $C^\infty(Z)$, and $G_{ab}(y)$ is a (pseudo-) Riemannian metric on $Y$. Actually, a \emph{semi-conformally flat noncommutative structure} on noncommutative manifold $X=Y\times Z$ is a pseudo-Riemannian metric $g_{\mu \nu}(y,z)=g_{ab}(y,z) \oplus h_{ij}(z)$ so that $h_{ij}(z)$ is the flat Kahler-Yau metric on $Z$, and $g_{ab}(y,z)=e^{\sigma(z)}G_{ab}(y)$, for $G_{ab}(y)$ an arbitrary (pseudo-) Riemannian metric on $Y$. In general, $X$ is considered to be Riemannian, but however, for significance of its applications in physics it can be assumed to admit the Lorentz signature $(-,+,\cdots,+)$, whilst the minus sign belongs to $g_{ab}$. Moreover, $h_{ij}$s are constant in the mentioned Darboux's charts due to the structure of the Kahler form: $h_{ij}=-\theta_{ik}J^k_j$, wherein $J^i_j$ is the almost complex structure.

\par Let $\nabla^X$ be the Levi-Civita connections for $g_{\mu \nu}=g_{ab}\oplus h_{ij}$. It is seen that $\Gamma^a_{ij}=\Gamma^i_{aj}=\Gamma^i_{ja}=\Gamma^i_{jk}=0$, $\Gamma^a_{ib}= \frac{\partial_i \sigma}{2} \delta^a_b$, and $\Gamma^i_{ab}=-\frac{h^{ij}g_{ab}}{2}\partial_j \sigma$. Also, the Christoffel symbols on $Y$ are given for the metric $G_{ab}$; $\Gamma^a_{bc}=\frac{G^{ad}}{2}(G_{bd,c}+G_{dc,b}-G_{bc,d})$. Hence;
\begin{equation} \label {khodash}
\nabla_i^X \text{d}x^j=0,~~~~~~~~~~\text{and;}~~~~~~~~~~\nabla_i^X \text{d}x^a \propto \text{d}x^a.
\end{equation}

\noindent All in all, we find:
\begin{equation} \label {flat}
[\nabla^X_i,\nabla^X_j]=0,~~~~~\text{or;}~~~~~{R^X_{ij \mu}}^\nu=0.
\end{equation}

\noindent We also read; ${R^X_{aij}}^k={R^X_{abi}}^j={R^X_{abi}}^c=0$, while the other components are generally nontrivial. Specifically,
\begin{equation} \label {Rx-Ry}
\begin{split}
{R^X_{abc}}^d={R^Y_{abc}}^d+\frac{1}{4}\left(\delta^d_b g_{ac} - \delta^d_a g_{bc} \right) \partial_i \sigma \partial^i \sigma,\\
{R^X_{aij}}^b=-\frac{\delta^b_a}{2} \left(\partial_i \partial_j \sigma - \frac{1}{2} \partial_i \sigma \partial_j \sigma \right),~~~~~~~~~~ \\
{R^X_{aib}}^j=\frac{g_{ab}}{2} \left( \partial_i \partial^j \sigma - \frac{1}{2} \partial_i \sigma \partial^j \sigma \right),~~~~~~~~~~~
\end{split}
\end{equation}

\noindent wherein ${R^Y_{abc}}^d$ is the curvature of the Levi-Civita connection $\nabla^Y$ of $g_{ab}$. In principle, the nontrivial topological features of noncommutative gravity for a semi-conformally flat noncommutative structure are subject to a Weyl rescaling on $Z$. One can easily see that $R^X_{ab}=R^Y_{ab}-\frac{g_{ab}}{2}\left(\frac{1}{2}\partial^i \sigma \partial_i \sigma-\partial^i \partial_i \sigma \right)$, $R^X_{ai}=0$, and $R^X_{ij}=-n\left( \partial_i \partial_j \sigma + \frac{1}{2} \partial_i \sigma \partial_j \sigma \right)$, for $R^X_{\mu \nu}$ (resp. $R^Y_{ab}$) the Ricci tensor of connection $\nabla^X$ (resp. $\nabla^Y$). In addition, we obtain $\mathcal R^X=\mathcal R^Y-n \partial^i \partial_i \sigma$, where $\mathcal R^X$ (resp. $\mathcal R^Y_{ab}$) is the Ricci scalar of $\nabla^X$ (resp. $\nabla^Y$). Thus, even before introducing a star product on $X$, the ordinary Einstein-Hilbert action of the vacuum is affected by the extra dimensions of $Z$ with a Klein-Gordon theory. However, this scalar field would not give rise to nontrivial effects for commutative fields, since it behaves like some Weyl rescaling. Indeed, via the mentioned formalism the nontrivial features of the Einstein-Hilbert theory arise only when one considers the contributions of the matter fields and their interactions with some noncommutative star product.

\par Now, we are ready to put a noncommutative structure on $X$ by defining a covariant noncommutative structure for general vector bundles over $X$. Set $\mathbb F=\mathbb R$ or $\mathbb C$ and let $\mathbb F^{K_i} \to E_i \to X$, $i=1,2,3$, be three arbitrary vector bundles endowed with a smooth bundle structure as; $\frak m:E_1 \otimes E_2 \to E_3$. Actually, $\frak m$ induces a multiplication of smooth sections, also denoted with the same symbol $\frak m$ as; $\frak m:\Gamma(E_1) \otimes \Gamma(E_2) \to \Gamma(E_3)$. To produce the most general formulation of star product we quantize the multiplication map $\frak m$ in a covariant framework. In principle, to quantize $\frak m$ consistently we should follow the well-understood methods of deformation quantization explained in \cite{bayen1, bayen2} and define a generalized Moyal-Weyl-Wigner quantization on $X$ with employing the connections on $E_i$s. Let $\nabla^i$ be a connection on $E_i$ and assume that $\hat \omega^{\sharp}=\theta^{ij} \partial_i \otimes \partial_j$ be the Poisson bivector as the dual of the Kahler form $\hat \omega$ obtained via converting the noncommutativity matrix $\theta_{ij}$, i.e. $\theta^{ij}\theta_{jk}=\delta^i_k$. We introduce a bilinear differential operator on $\Gamma(E_1) \otimes \Gamma(E_2)$ with;
\begin{equation} \label {NCSM}
\Theta=\theta^{i j }\nabla^1_i ~\otimes \nabla^2_j,
\end{equation}

\noindent which is the contraction of $\hat \omega^{\sharp}$ and $\nabla^1 \otimes \nabla^2=\nabla^1_\mu \text{d}x^\mu \otimes \nabla^2_\nu \text{d}x^\nu$. Obviously $\Theta$ is globally well-defined on $X$. Now, we introduce the \emph{covariant star product} $\star$ as a generalized Groenewold-Moyal star product for smooth sections $\omega_i \in \Gamma(E_i)$, $i=1,2$, as:
\begin{equation} \label {star}
\begin{split}
\omega_1 \star \omega_2:=\frak m  \left( e^{i\Theta } \left(\omega_1 \otimes \omega_2\right) \right)=\frak m \left( e^{i\theta^{ij} \nabla^1_i \otimes \nabla^2_j} \left(\omega_1 \otimes \omega_2\right) \right)~~~~~~~~~~~~~~~~~\\
=\sum_{N\geq 0} \frac{i^N}{N!} \theta^{i_1 j_1} \cdots \theta^{i_N j_N} \frak m\left( \left(\nabla^1_{i_N} \cdots \nabla^1_{i_1} \omega_1\right) \otimes  \left(\nabla^2_{j_N} \cdots \nabla^2_{j_1} \omega_2 \right) \right)\\
=\sum_{N\geq 0}\sum_{~~|(i)|,|(j)|=N} \frac{i^N}{N!} \theta^{(i) (j)} \frak m\left( \left(\nabla^1_{(i)} \omega_1\right) \otimes  \left(\nabla^2_{(j)} \omega_2 \right) \right),~~~~~~~~~
\end{split}
\end{equation}

\noindent wherein $(i)$ is a short notation for multi-indices $(i_1, \cdots, i_N)$ with degree $|(i)|=N$, and $\nabla_{(i)}$ stands for multiple covariant derivatives $\nabla_{i_N} \cdots \nabla_{i_1}$. Also, we assume $\nabla_{(i)}\omega=\omega$ for $N=0$. As we read from (\ref {star}) the covariant star product after the second order of $\theta$ is essentially rescaled by the curvature of connections $\nabla^i$s. This is the geometric realization of dynamical noncommutative spacetime for noncommutative gravity.\footnote{See \cite{steinacker} for more discussions.}

\par It is worth to note that the star product (\ref {star}) is smoothly defined everywhere on $X$, and for the flat connection $\nabla^i_\mu=\partial_\mu$, $i=1,2,3$, on trivial vector bundles $E_i=X \times \mathbb F^{K_i} $, the covariant star product $\star$ naturally leads to the well-known Groenewold-Moyal product $\star_{G-M}$. Indeed, this situation arises for example when the base manifold is contractible, e.g. $X=\mathbb R^4$, which appears in quantum field theory. Specially, when $E_1=E_2=E_3=X \times \mathbb R$, one finds: $\Gamma(E_i)=C^\infty(X)$, $i=1,2,3$, thus we would have; $\Theta=\theta^{\mu \nu} \partial_\mu \otimes \partial_\nu$, which readily results in: $\star=\star_{G-M}$. In this case $\star$ becomes an associative product. In fact, if $E_1=E_2=E_3=E$ is trivial and $\nabla^1_\mu=\nabla^2_\mu=\nabla^3_\mu=\partial_\mu$, the covariant star product $\star$ would be an associative structure on $\Gamma(E)$, whenever $\frak m$ is associative too.

\par When $E_1=E_2=E_3=\oplus_{k=0}^{2m}\Lambda^k T^*X$ the mentioned quantization method is accomplished with differential forms on $X$ for $\Gamma(E_i)=\Omega^*(X)$, $i=1,2,3$, whereas $\frak m$ is the ordinary wedge product. In this case, as we consider $\nabla^i=\nabla^X$ for each $i$, then the star product $\star$ would be still an associative multiplication; i.e. $\omega_1 \star (\omega_2 \star \omega_3)=(\omega_1 \star \omega_2 ) \star \omega_3$. This is in fact, the immediate consequence of the Baker-Campbell-Hausdorff formula and (\ref {flat}). Moreover, for any arbitrary topology and connection of $E\to X$ the scalar multiplication is always associative, i.e. for each $f,g \in C^\infty(X)$ and $\zeta \in \Gamma(E)$, we read: $f\star (g \star \zeta)=(f \star g) \star \zeta$. Indeed, it is also an apparent result of (\ref {star}).

\par Substantially, if $E_1=E_2=E_3=E$ is not nontrivial, the covariant star product $\star:\Gamma(E)\otimes \Gamma(E) \to \Gamma(E)$ is not generally associative, even for associative $\frak m$. However, if $E\to X$ admits a connection $\nabla$ with vanishin curvature on $Z$, then $\star$ would be associative too. In fact, if $[\nabla_i,\nabla_j]=0$, then upon the Baker-Campbell-Hausdorff formula the covariant star product $\star$ as the quantization of $\frak m$ would be still associative. Such vector bundles (resp. connections) are referred to as \emph{associative vector bundles} (resp. \emph{associative connections}). Indeed, whenever $E$ is associative we consider an associative connection in quantizing $\frak m$.

\par Actually, the star product quantization of differential structures over $X$ could be consistently implemented by means of (\ref {star}). Also, the inner products on vector bundles could be quantized via the covariant star product (\ref {star}). Let $\mathbb F^K \to E\to X$ be a vector bundle endowed with a fiberwise smooth metric $\frak g_{pq}$, defined locally for the local frame $\{e_p \}$. This induces a multiplication as;
\begin{equation} \label {inner}
\frak g:\Gamma(E) \otimes \Gamma(E)\to C^\infty(X),~~~~~~~~~~\text{with;}~~~~~~~~~~\frak g (\zeta^p e_p \otimes \xi^q e_q)=\frak g_{pq}\zeta^p \xi^q.
\end{equation}

\noindent Therefore, upon (\ref {star}) the quantization of $\frak g$, as a covariant star product is locally given as:
\begin{equation} \label{inner-star}
\zeta \star \xi=\sum_{N\geq 0} \sum_{~~|(i)|,|(j)|=N} \frac{i^N }{N!} \theta^{(i)(j)}  \frak g_{pq} \left( \nabla_{(i)} \zeta^p \right)\left( \nabla_{(j)} \xi^q \right).
\end{equation}

\noindent where $\nabla$ is the connection of $E$. Specially, if $\mathcal X$ and $\mathcal X'$ are two vector fields on $X$, then their star inner product is:
\begin{equation} \label {xx'}
\mathcal X \star \mathcal X'=\sum_{N\geq 0} \sum_{~~|(i)|,|(j)|=N}  \frac{i^N }{N!} \theta^{(i)(j)}  g_{\mu \nu} \left( \nabla^X_{(i)} \mathcal X^\mu \right)\left( \nabla^X_{(j)} \mathcal X'^\nu \right).
\end{equation} 

\noindent It can be seen that (\ref {inner-star}) is independent of the chosen local frame $\{e_p\}$ for $E$, and hence is a well-defined structure. To see this let $\{ e'_p\}$ be another local frame for $E$ defined over some trivialization chart. Then, for any smooth section $\zeta \in \Gamma(E)$ we have; $\zeta'^p=U^p_q \zeta^q$, for some invertible matrix $U$. On the other hand, in the basis of $\{ e'_p \}$ the connection $\nabla$ becomes; $\nabla'=U \nabla U^{-1}$. Thus, we read: $\nabla'_{(i)}\zeta'^p=U^p_q \nabla_{(i)} \zeta^q$. Moreover, in the basis $\{e'_p\}$ the fiber metric is: $\frak g'_{pq}={U^{-1}}_p^r \frak g_{rs} {U^{-1}}^s_q$. Consequently;
\begin{equation} \label {inner-well-defined}
\frak g'_{pq} \left( \nabla'_{(i)}  \zeta'^p  \right) \left(\nabla'_{(i)}\xi'^q \right)=\frak g_{pq}\left( \nabla_{(i)} \zeta^p \right) \left( \nabla_{(i)}\xi^q \right),
\end{equation}

\noindent which confirms the consistency of (\ref {inner-star}). Similarly, we see that if $E$ is a complex vector bundle endowed with a Hermitian metric $\frak h: \Gamma(E) \otimes \Gamma(E) \to C^\infty(X)$, and the connection $\nabla$, the quantization of $\frak h$ via
\begin{equation} \label {hermit-star}
\zeta \star \xi=\sum_{N\geq 0} \sum_{~~|(i)|,|(j)|=N}  \frac{i^N }{N!} \theta^{(i)(j)}  \frak h_{pq} \left( \nabla_{(i)} \zeta^p \right) \overline{ \left( \nabla_{(j)} {\xi}^q \right)}
\end{equation}

\noindent is well-defined and independent of the chosen local frame. However, as we see from (\ref{inner-star}) the fiber metric components $\frak g_{pq}$s would not be covariantly differentiated in the process of multiplying two sections with $\star$. This is, in principle, the main reason for the consistency of (\ref {inner-star}). Nonetheless, the star inner product (\ref {inner-star}) is not bilinear for the scalar star product; i.e. if $f\in C^\infty(X)$ and $\zeta, \xi \in \Gamma(E)$, then $f \star (\zeta \star \xi)$, $(f \star \zeta) \star \xi$, and $\zeta \star (f \star \xi)$ are generally different. 

\section{Some Aspects of Noncommutative Physics on Semi-Conformally Flat Noncommutative Spacetime with Covariant Star Product}
\setcounter{equation}{0}

\par Obviously, although the star inner product is well-defined it is not a local structure. However, for non-smooth objects on $X$ we naively assume the covariant derivative to vanish identically. For instance, for two tangent vectors $\mathcal V, \mathcal V' \in T_pX$, we readily suppose: $\nabla^X \mathcal V=\nabla^X \mathcal V'=0$, hence; $\mathcal V \star \mathcal V'=\mathcal V. \mathcal V'$. Also, we obtain $\nabla^X_{\mathcal V} \mathcal X=0$, whenever $\mathcal X$ is a vector field along a smooth curve in $X$, and $\mathcal V$ is a tangent vector not proportional to $\mathcal X$. Therefore, $\star$ will not appear in the geodesic equations on $X$. In principle, a geodesic on $X$ is a solution of equations:
\begin{equation} \label {geod}
\begin{split}
\frac{\text{d}^2 x^a}{\text{d} t^2}+\Gamma^a_{bc}\frac{\text{d} x^b}{\text{d}t} \frac{\text{d} x^c}{\text{d}t}+ \partial_i \sigma \frac{\text{d} x^a}{\text{d}t} \frac{\text{d} x^i}{\text{d}t}=0,\\
\frac{\text{d}^2 x^i}{\text{d} t^2}-\frac{ \partial^i \sigma}{2} g_{ab} \frac{\text{d} x^a}{\text{d}t} \frac{\text{d} x^b}{\text{d}t}=0.~~~~~~~~~~~~~~
\end{split}
\end{equation}

\noindent Actually, from the viewpoint of general relativity the first equation of (\ref {geod}) implies that the observable trajectory of an inertial body is affected by the Weyl rescaling factor of the extra dimensions of $Z$ and their corresponding (unobservable) components of the velocity. On the other hand, the second equation of (\ref {geod}) states that the extra dimensional components of the velocity could not be vanished unless $\frac{\text{d}\gamma}{\text{d}t}=0$ on $Y$, i.e. for light like geodesics in $Y$ whenever $g_{ab}$ is Lorentzian. Hence, although $\star$ does not take part in geodesic equations, the geometries of the geodesics are affected by the extra dimensions of $Z$ for massive particles. More precisely, as far as we use semi-conformally flat noncommutative structures the noncommutative effects of gravity would appear only in presence of matter fields.

\par The covariant star product $\star$ on differential structures of $X$ has some other interesting features with significance in noncommutative gravity. For instance, since $\nabla^X g_{\mu \nu}=0$, we easily see that the contraction of any arbitrary tensor field with $g_{\mu \nu}$ receives no contribution from the noncommutativity. Specially, rising and lowering the indices of tensors taken place ordinarily on noncommutative manifold $X$. Therefore, the determinant of the metric $g_{\mu \nu}$ is calculated via the usual way in presence of $\star$, i.e.;
\begin{equation} \label {determinant2}
g_\star=\frac{1}{(2n)!}\epsilon^{\mu_1 \cdots \mu_{2n}} \epsilon^{\nu_1 \cdots \nu_{2n}}g_{\mu_1 \nu_1} \star \cdots \star g_{\mu_{2n} \nu_{2n}}=\frac{1}{(2n)!}\epsilon^{\mu_1 \cdots \mu_{2n}} \epsilon^{\nu_1 \cdots \nu_{2n}}g_{\mu_1 \nu_1}  \cdots  g_{\mu_{2n} \nu_{2n}}=g.
\end{equation}

\noindent Hence, the volume form of $X$ is unchanged as we transfer from ordinary to noncommutative structures on $X$, i.e. $\text{dvol}_\star=\text{dvol}$. Meanwhile, the Christoffel symbols remain invariant via transferring from the ordinary product to the covariant star product $\star$; $\Gamma_{\star \mu \nu}^\sigma=\Gamma^\sigma_{\mu \nu}$. This, confirms the mentioned claims about the geodesic equations on $X$. Nonetheless, the noncommutative Riemann curvature tensor ${R^X_{\star \mu \nu \sigma}}^\lambda$, produced by employing $\star$, will not coincide with its ordinary version ${R^X_{\mu \nu \sigma}}^\lambda$ generally. The main difference here is that the commutator of Christoffel symbols via the star product $\star$ would add nontrivial terms to ${R^X_{\star \mu \nu \sigma}}^\lambda$. However, the structure of $\star$ is so that ${R^X_{\star \mu \nu \sigma}}^\lambda$ remains well-defined and gauge (general coordinate transformation) covariance. Moreover, the noncommutative Ricci tensor $R^X_{\star \mu \nu}={R^X_{\star \sigma \mu \nu}}^\sigma$ and the noncommutative Ricci scalar $\mathcal R_\star^X= R^X_{\star \mu \nu}g^{\mu \nu}$ both have only one star product included. 

\par In order to compare the noncommutative version of the Einstein-Hilbert action with its ordinary formula we need to prove an important lemma. However, let us give some preliminaries first: Suppose that the vector bundle $E \to X$ and the connection $\nabla$ are associative. We also assume that $E$ admits a fiberwise metric $\frak g_{pq}$, the so called \emph{associative metric}, which is compatible with $\nabla$, i.e. $\nabla_i \frak g_{pq}=0$ for each $D \leq i <2n$. Thus, we can simply write (\ref {inner-star}) as;
\begin{equation} \label {inner-associative}
\zeta \star \xi=\sum_{N\geq 0} \sum_{~~|(i)|,|(j)|=N} \frac{i^N }{N!} \theta^{(i)(j)}  \left( \nabla_{(i)} \zeta_p \right)\left( \nabla_{(j)} \xi^p \right),
\end{equation}

\noindent for $\zeta_p= \frak g_{pq} \zeta^q$. In principle, from now on for all associative vector bundles, we would assume associative connections and metrics accordingly. Now, we can prove an important lemma that has a significant role in our study.\\

\par \textbf{Lemma 1;} \emph{Assume that $E\to X$ is associative. Then;}
\begin{equation}
~~~~~~~~~~~~~~~\int_X *(\zeta \star \xi) =\int_X *(\zeta.\xi),~~~~~~~~~~~~~~~\zeta, \xi \in \Gamma(E),
\end{equation}

\noindent \emph{for $*$ the Hodge star operator and for $\zeta.\xi$ the ordinary inner product of $\zeta$ and $\xi$ in $\Gamma(E)$.}\\

\par \textbf{Proof;} At first we should note that if $\mathcal X$ is a vector field on $X$, then:
\begin{equation} \label {divergence}
*(\nabla^X_\mu \mathcal X^\mu)=\text{d}(\frak i_{\mathcal X} \text{dvol}),
\end{equation}

\noindent for $\frak i_{\mathcal X}$ the interior product with the vector field $\mathcal X$. We should recall that $\partial_i$s are globally defined vector fields on $X$. Therefore, if $\mathcal X=\mathcal X^i \partial_i$ is a vector field on $X$ we have;
\begin{equation} \label {d}
* (\nabla^X_i \mathcal X^i )=\text{d} \left( \mathcal X^i (\frak i_{\partial_i} \text{dvol}) \right).
\end{equation}

\noindent Moreover, apart from the leading term the inner star product $\zeta \star \xi$ consists of the following terms;
\begin{equation} \label {d-2}
\theta^{ij} \left( \nabla_i \zeta'_p \right) \left( \nabla_j \xi'^p \right) =\partial_i \left(\theta^{ij} \zeta'_p \nabla_j \xi'^p \right)-\frac{\theta^{ij}}{2} \left( \zeta'_p \right) \left( [ \nabla_i ,\nabla_j ] \xi'^p \right)=\nabla^X_i \left(\theta^{ij} \zeta'_p \partial_j \xi'^p \right).
\end{equation}

\noindent Hence, if we set $\mathcal X^i=\theta^{ij} \zeta'_p \nabla_j \xi'^p$, then (\ref {d}) and (\ref {d-2}) lead to;
\begin{equation} \label {d-2'}
*(\zeta \star \xi)=*(\zeta. \xi) + \text{d}\beta,
\end{equation}

\noindent for $\text{d}\beta$ an exact form on $X$. Thus, the lemma follows. \textbf{Q.E.D.}\\

\par The above lemma leads to two important properties of $\star$ of $C^\infty(X)$ which are in common with the Groenewold-Moyal star product and have significant applications in noncommutative gravity: The \textbf{Triviality} and the \textbf{Traciality};\\

\par \textbf{a) Triviality;} \emph{If $f_1, f_2 \in C^\infty(X)$, then; $* (f_1 \star f_2)= * (f_1 f_{2} )+ \text{d}\beta$, for $\emph{d}\beta$ an exact form on $X$. Hence, if $f_1, f_2 \in C^\infty(X)$, then;}
\begin{equation} \label {triviality}
\int_X *(f_1 \star f_2) =\int_X  *(f_1 f_2).
\end{equation}

\noindent \emph{In principle, $*(f_1 \star f_2)$ and $*(f_1 f_2)$ belong to the same cohomology class in $H^{2n}_{\emph {dR}} (X,\mathbb{R})$.}

\par \textbf{b) Traciality;} \emph{If $\{f_i \}_{i=1}^s$ is a set of smooth functions on $X$, then;}
\begin{equation} \label {traciality}
*( f_1 \star \cdots \star f_s)= *(f_s \star f_1 \star \cdots \star f_{s-1}) + \text{d}\beta.
\end{equation}

\noindent \emph{Hence, if $f_1, f_2, f_3\in C^\infty(X)$, then;}
\begin{equation}
\int_X *(f_1 \star f_2 \star f_3) =\int_X *(f_3 \star f_1\star f_2).
\end{equation}
~

\par Therefore, upon the statement of \textbf{Triviality}, the noncommutative Einstein-Hilbert action has no $\star$ included, although it contains a Weyl rescaling factor $e^\sigma(z)\in C^\infty(Z)$ of the (unobservable) noncommutative part of spacetime. Actually, we have;
\begin{equation} \label {einstein}
S_{\star E-H}=\frac{1}{16\pi G} \int_X \mathcal *R^X_\star =\frac{1}{16\pi G} \int_X  * \mathcal R^X =S_{E-H},
\end{equation}

\noindent wherein we used the identity $f\star \text{dvol}=f \text{dvol}$, for any $f\in C^\infty(X)$. In principle, the vacuume solution of Einstein's field equation on a noncommutative spacetime, receives no contribution from covariant star product $\star$. As a special case the Schwarzchild black hole has exactly the same gravitational features as its noncommutative version provided a semi-conformally flat noncommutative structure is considered on $X$. But, however, when the matter fields are included, the noncommutative gravitational features will appear accordingly. Thus, it seems that upon the semi-conformally flat noncommutative structures the noncommutativity of spacetime would be a dynamical entity; the idea that was noted in \cite{steinacker}.\footnote{See also \cite{nicolini} for more discussions.}

\par It is worth to note that if $Y$ is a $4$-manifold, then the black hole solution on some contractible $U^X_\alpha$ is simply given by considering $G_{ab}$ to be the Schwarzschild metric, i.e.;
\begin{equation} \label {black hole}
\text{d}s^2=e^{\sigma(x^4,\cdots, x^{2n-1})} \left( -\left( 1- \frac{r_s}{r_\alpha}\right) c^2 \text{d}t_\alpha^2 + \left( 1- \frac{r_s}{r_\alpha} \right)^{-1} \text{d}r_\alpha^2 + r_\alpha^2 \left( \text{d}\theta_\alpha^2 + \sin^2 \theta_\alpha \text{d} \phi_\alpha^2 \right) \right)+\sum_i {\text{d}x^i_\alpha}^2,
\end{equation}

\noindent wherein $t_\alpha=x^0_\alpha$, and $(r_\alpha,\theta_\alpha, \phi_\alpha)$ is the spherical coordinate on the space like submanifold of $U^Y_\alpha$. However, the Weil rescaling power $\sigma \in C^\infty(Z)$ is the solution of the following Euler-Lagrange equation:
\begin{equation} \label {power}
\partial_i \partial^i \sigma - \frac{1}{2n} \mathcal R^Y + \frac{D}{4} \partial_i \sigma \partial^i \sigma=0.
\end{equation}

\noindent Let $Z$ be $2m$-dimensional torus. Then, the Fourier transformation of (\ref {power}) is;
\begin{equation} \label {fourier-sigma}
p^2a_p+\frac{1}{2n}R^Y\delta_{p,0}+\frac{D}{4}(p-q)_iq^ia_{p-q}a_q=0,
\end{equation}

\noindent with $a_p$ the Fourier coefficient of $\sigma$ for the Fourier mode $p=(p_1,\cdots,p_{2m})\in \mathbb Z^{2m}$. The solution for $p=0$ leads to $|a_q|^2=\frac{\mathcal R^Y}{2n q^2}$, $q \ne 0$, which yield the following equation;
\begin{equation} \label {R>0}
\mathcal R^Y \geq 0.
\end{equation}

\noindent Hence, for the vacuum spacetime of the Schwarzchild black hole the classical solution results in constant $\sigma$. However, in presence of matter fields, $\sigma$ cannot be trivial and gives rise to a nontrivial topological constraint on $Y$ via (\ref {R>0}) as an obstruction for admitting negative Ricci scalar. Actually, $\sigma$ can be compared with gravitational dilatons such as the Kaluza-Klein radion. Therefore, if $Y$ is a $4p$-dimensional, then upon the Lichnerowicz obstruction theorem \cite{lic}, in order to fulfill (\ref {R>0}) the de Rham cohomology class of $\hat A_p(TY)$ must be vanished identically, where $\hat A_p(TY)$ is the $p$-th Dirac genus of $TY$. 

\par Another immediate consequence of \textbf{Triviality} is that the kinetic action of noncommutative scalar field theory in presence of gravity coincides exactly with that of the ordinary commutative case. In fact, the noncommutative Klein-Gordon theory includes no $\star$, i.e.;
\begin{equation} \label {NC-K-G}
S_{\star K-G} =\frac{1}{2} \int_X *(\text{d} \phi \star \text{d} \phi + m^2 \phi \star \phi )=\frac{1}{2} \int_X *( \text{d} \phi . \text{d} \phi + m^2 \phi^2)=S_{K-G},
\end{equation}

\noindent wherein the equality of $\text{d}\phi . \text{d} \phi=g^{\mu\nu} \partial_\mu \phi \partial_\nu \phi=\partial_\mu \phi \partial^\mu \phi$, is the inner product in associative bundle $T^*X$. Hence, even in presence of a background noncommutative gravity, the quantum effects are affected by $\star$ only for the interaction of matter fields. Therefore, the expectation value of the effective noncommutativity matrix must be substantially a dynamical entity. This is in fact, an immediate consequence of the semi-conformally flat noncommutative structure and the consistent construction of covariant star product. 

\par It can be easily seen that the kinetic terms of noncommutative gauge theories, including the Yang-Mills and the Maxwell theory, are both out of $\star$. Actually, if $A$ is a local Yang-Mills or Abelian gauge field, then by \textbf{Lemma 1} the kinetic term of the corresponding noncommutative gauge theory is locally given by $*(\text{d}A \star \text{d}A)$ which equals with $*(\text{d}A. \text{d}A)$ up to some exact term. Here, $\star$ is the quantization of the inner product of two-forms in $\Omega^2(X)$ given as: $\omega.\omega'=\omega_{\mu \nu} \omega'^{\mu \nu}$. Actually, we read: $\text{d}A. \text{d}A=(\partial_\mu A_\nu - \partial_\nu A_\mu)(\partial^\mu A^\nu-\partial^\nu A^\mu)$.

\par We conclude this section with a notable remark about $\star$. Indeed, it is easy to see that the covariant star product $\star$ provides an involutive structure via complex conjugation. Particularly, since the Kahler form $\hat \omega$ is real, then $\star$ behaves similar to the Groenewold-Moyal product versus complex conjugation, i.e. if $\omega_i \in \Omega^{n_i}(X)$, with $i=1,2$, then;
\begin{equation} \label{star-conj}
\overline{\omega_1 \star \omega_2}=(-1)^{n_1 n_2} \overline{\omega_2} \star \overline{\omega_1}.
\end{equation}

\noindent Therefore, the noncommutative Dirac action would be real and well-defined whenever a semi-conformally flat noncommutative structure is assumed on $X$. Indeed, similarly, we can see that the kinetic term of the Dirac action contains no $\star$.


\section{Covariant Star Product, Noncommutative Characteristic Classes and Noncommutative Chern-Weil Theorem}
\setcounter{equation}{0}

\par Assume that $\mathbb F^{K_i} \to E_i \to X$, $i=1,2$, are vector bundles and $\Psi:\Gamma(E_1) \to \Gamma(E_2)$ be a pseudo-differential linear operator of order $r$. Therefore, $\Psi$ induces a vector bundle homomorphism as; $\Psi:\frak J^r(E_1) \to E_2$, wherein $\frak J^r(E_1)$ is the $r$-th jet bundle of $E_1$. In other words, $\Psi$ is a section of vector bundle $\mathbb F^K \to \Xi=E_2 \otimes \frak J^r(E_1)^*\to X$, with $K=K_1K_2 {2n+r \choose r}$. Actually, $\Xi$ is endowed with the induced connection $\nabla^\Xi=\nabla^2\otimes 1+1 \otimes \nabla^1$. The noncommutative version of $\Psi$, denoted with $\Psi_\star$, is defined by quantizing the following map:
\begin{equation} \label {D-1}
\frak m:\Gamma(\Xi) \otimes \Gamma(E_1) \to \Gamma(E_2). 
\end{equation}

\noindent Hence, if $\zeta \in \Gamma(E_1)$, then we obtain:
\begin{equation} \label {D-2}
\Psi_\star(\zeta)=\sum_{N \geq 0} \sum_{~~|(i)|,|(j)|=N}  \frac{i^N \theta^{(i)(j)}}{N!}  \nabla^\Xi_{(i)} \Psi  \left( \nabla^1_{(j)} \zeta \right),
\end{equation}

\noindent with $\nabla^\Xi_{(i)} \Psi=\Psi$ for $N=0$, $\nabla^\Xi_{(i)} \Psi=\nabla^2_i \Psi -\Psi \nabla^1_i$ for $N=1$, and
\begin{equation}
\nabla^\Xi_{(i)} \Psi=\nabla^2_{i_N} \left( \nabla^\Xi_{(i')} \Psi \right)-  \left( \nabla^\Xi_{(i')} \Psi \right) \nabla^1_{i_N},
\end{equation}

\noindent for $N >1$, where $(i)=(i_1, \cdots, i_N)$ and $(i')=(i_1, \cdots, i_{N-1})$. Specially, when $E_1=E_2=E$, and $\nabla^1=\nabla^2=\nabla$, we have;
\begin{equation} \label {D-3}
\Psi_\star(\zeta)
=\sum_{N \geq 0} \sum_{~~|(i)|,|(j)|=N} \frac{i^N \theta^{(i)(j)}}{N!} [ \nabla_{(i)} , \Psi] \left( \nabla_{(j)} \zeta \right).
\end{equation}

\noindent Also, the quantization of composition of two pseudo-differential operators is thus given by (\ref {star}) as;
\begin{equation} \label {composition}
\Psi_1 \star \Psi_2
=\sum_{N \geq 0} \sum_{~~|(i)|,|(j)|=N} \frac{i^N \theta^{(i)(j)}}{N!} [ \nabla_{(i)} , \Psi_1 ] \circ [\nabla_{(j)}, \Psi_2 ].
\end{equation}

\noindent Hence, if $E$ is associative and $\Psi_1, \Psi_2:\Gamma(E)\to \Gamma(E)$ are two pseudo-differential operators, then:
\begin{equation} \label {2 op}
~~~~~~~~~~\Psi_{1 \star} \left(\Psi_{2 \star} (\zeta) \right)= \left( \Psi_1 \star \Psi_2 \right)_\star (\zeta),~~~~~~~~~~\zeta \in \Gamma(E).
\end{equation}

\par In principle, we obtain:
\begin{equation} \label {nabla-nabla}
\begin{split}
[\nabla_\mu , \nabla_\nu ]_\star= \nabla_\mu \star \nabla_\nu - \nabla_\nu \star \nabla_\mu  ~~~~~~~~~~~~~~~~~  \\
=\sum_{N \geq 0} \sum_{~~|(i)|,|(j)|=2N} \frac{(-1)^N \theta^{(i)(j)}}{(2N)!} \left[ [ \nabla_{(i)} , \nabla_\mu] , [\nabla_{(j)}, \nabla_\nu] \right] ~~~~ \\
+ i \sum_{N \geq 0} \sum_{~~|(i)|,|(j)|=2N+1} \frac{(-1)^N \theta^{(i)(j)}}{(2N+1)!} \{ [ \nabla_{(i)} , \nabla_\mu] , [\nabla_{(j)}, \nabla_\nu] \}.
\end{split}
\end{equation}

\noindent The noncommutative curvature tensor is defined by coupling $[\nabla_\mu,\nabla_\nu]_\star$ with $\text{d}x^\mu \wedge \text{d}x^\nu$ as $F_{\star}=F_{\star \mu \nu} \text{d} x^{\mu} \wedge \text{d}x^\nu$. Therefore, the anti-commutator terms of (\ref {nabla-nabla}) would be removed and $F_\star$ is given as;
\begin{equation} \label {f-star cov}
F_{\star \mu \nu}=\sum_{N \geq 0} \sum_{~~|(i)|,|(j)|=2N} \frac{(-1)^N \theta^{(i)(j)}}{(2N)!} \left[ [ \nabla_{(i)} , \nabla_\mu] , [\nabla_{(j)}, \nabla_\nu] \right].
\end{equation}

\noindent Actually, we obtain:
\begin{equation} \label {f-star formula}
\begin{split}
F_{\star \mu \nu} =F_{\mu \nu} - \frac{1}{2!} \theta^{i_1 j_1} \theta^{i_2 j_2} [[\nabla_{i_1},F_{i_2 \mu}],[\nabla_{j_1}, F_{j_2 \nu} ]] ~~~~~~~~~~~~~~~~~\\
  + \frac{1}{4!} \theta^{i_1 i_1} \theta^{i_2 j_2}\theta^{i_3 j_3} \theta^{ i_4 j_4} [[\nabla_{i_1},[\nabla_{i_2},[\nabla_{i_3},F_{i_4 \mu}]]],[\nabla_{j_1},[\nabla_{j_2},[\nabla_{j_3}, F_{j_4 \nu} ]]]]+ \cdots.
\end{split}
\end{equation}

\noindent In principle, since $F_\star$ contains no anti-commutator, it would be an element of the Lie algebra of the structure group. In other words, if $G$ is the structure group of $\mathbb F^K \to E \to X$ with connection $\nabla$, then $F_\star \in \Omega^2(X) \otimes \frak g$, for $\frak g$ the Lie algebra of $G$. Moreover, if $E$ is an associative vector bundle then; $F_{\star ij}=F_{ij}=0$. But, however, there are more similarities between $F_{\star \mu \nu}$ and $F_{\mu \nu}$. The following lemma proves that $F_{\star \mu \nu}$ is a gauge covariant, and consequently a well-defined tensor.\\

\par \textbf{Lemma 2;} \emph{A change of local frame (i.e. a gauge transformation) in $E \to X$, given as}
$$\nabla=\text{d}+A ~~~~~~~~~~\to~~~~~~~~~~ \nabla'=U^{-1} \nabla U=\text{d}+U^{-1}AU+U^{-1}\text{d}U$$
\noindent \emph{leads to: $F'_{\star}=U^{-1} F_\star U$.}
~\\

\par \textbf{Proof;} If $A$ and $B$ are two pseudo-differential operators, then: $[U^{-1}AU,U^{-1}BU]=U^{-1}[A,B]U$. Therefore, since $\nabla' = U^{-1} \nabla U$ and $F'_{\mu \nu}=U^{-1} F_{\mu \nu} U$, by employing induction on consecutive terms of sequence (\ref {f-star cov}) (or (\ref {f-star formula})), it is seen that $F'_{\star \mu \nu}=U^{-1}F_{\mu \nu} U$. \textbf{Q.E.D.}\\

\par The following lemma establishes one of the most important identities for $F_\star$.\\

\par \textbf{Lemma 3;} \emph{The noncommutative curvature $F_\star$ fulfills the Bianchi identity, i.e.;}
\begin{equation}
\nabla_\mu F_{\star \nu \sigma}+\nabla_\nu F_{\star \sigma \mu} + \nabla_\sigma F_{\star \mu \nu}=0,
\end{equation}  

\noindent \emph{or; $[\hat \nabla , F_\star]_s=0$, wherein $\hat \nabla=\emph{d}x^\mu \wedge \nabla_\mu$, and $[,]_s$ is the supercommutator of differential forms.}\\

\par \textbf{Proof;} According to (\ref {f-star formula}) the noncommutative curvature $F_\star$ is expressed as;
\begin{equation} \label {t and t}
F_\star =\sum_{(i),(j)} \Phi^{(i)(j)} [T_{(i)\mu },T_{(j) \nu }]~\text{d}x^\mu \wedge \text{d}x^\nu,
\end{equation}

\noindent wherein $(i)$ and $(j)$ are multi-indices of even degrees, $T_{(i) \mu}$ is the commutator of $\nabla_{(i)}$ and $\nabla_\mu$, and $\Phi^{(i)(j)}$ is proportional to $\theta^{(i)(j)}$, hence a symmetric tensor. On the other hand, we have;
\begin{equation} 
\begin{split}
[\hat \nabla, [T_{(i) \mu},T_{(j)\nu}]]_s~ \Phi^{(i)(j)}~\text{d}x^\mu \wedge \text{d}x^\nu ~~~~~~~~~~~~~~~~~~~~~~~~\\
=\left( [[\nabla_\sigma,T_{(i) \mu}],T_{(j) \nu}]+[[\nabla_\sigma,T_{(j) \nu}],T_{(i) \mu}] \right)~ \Phi^{(i)(j)}~\text{d}x^\sigma \wedge \text{d}x^\mu \wedge \text{d}x^\nu,
\end{split}
\end{equation}

\noindent which vanishes identically, since it is a contraction of a symmetric and an antisymmetric tensor. This proves the noncommutative Bianchi identity. \textbf{Q.E.D.}\\

\par In principle, $F_\star$ is the most significant differential form for semi-conformally flat noncommutative manifolds. Actually, despite the naive noncommutative formula of the curvature $F_{\star_{G-M}}$, given by the standard form of Groenewold-Moyal product $\star_{G-M}$, the noncommutative curvature $F_\star$ introduced by (\ref {f-star formula}) gives rise to a numbers of interesting and well-defined differential geometric achievements. To see them, first of all one should note that $F_\star$, produced by covariant star product, in contrast to $F_{\star_{G-M}}$, is a well-defined smooth differential form on $X$. This is the main reason of its gauge covariance. On the other hand, this property helps us to create some well-defined cohomology classes by means of $F_\star$. Actually, if $\mathbb F^K \to E \to X$ is a vector bundle over $X$ with noncommutative curvature $F_\star$, and $P$ is a symmetric invariant polynomial on the Lie algebra $\frak {gl}(K,\mathbb F)=\mathbb M_{K\times K}(\mathbb F)$, then $P(F_\star)$ is a well-defined differential form on $X$ which as we will establish in below carries some significant pieces of geometric/topological information about $X$.

\par Actually, $P(F_\star)$ and its commutative version $P(F)$, provided by the ordinary curvature $F$, have several common significant properties. However, if $E$ is associative, then we can replace the matrix product in the polynomial $P$ with $\star$ due to (\ref {star}). Actually, we can see that this noncommutative version of $P(F_\star)$, denoted as $P_\star(F_\star)$, also produces a well-defined (i.e. gauge invariant) differential form. But, however, $P_\star(F_\star)$ is not closed in general, hence would not define a de Rham cohomology class. Therefore, we may use $P(F_\star)$ in studying the noncommutative characteristic classes. The next theorem is in fact, the noncommutative version of the celebrated Chern-Weil theorem. \\

\par \textbf{Theorem 1;} \emph{Assume that $\mathbb F^K \to E \to X$ is an associative vector bundle with connection $\nabla$ and $\star$ is its star product. Also, let $P$ be a symmetric invariant polynomial on $\mathbb M_{K \times K}(\mathbb F)$. Then, we have:}
\par \textbf{a)} \emph{$P(F_\star)$ is a closed differential form.}
\par \textbf{b)} \emph{The cohomology class of $P(F_\star)$ is independent of the connection $\nabla$.}
\par \textbf{c)} \emph{$P(F)$ and $P(F_\star)$ are cohomologous, i.e. $P(F)=P(F_\star)+\emph{d} \lambda$ for some exact form $\emph{d} \lambda$.}\\

\par \textbf{Proof;} \textbf{a)} Let $P$ be of degree $r$. Then, since $P$ is symmetric and invariant it is seen that;
\begin{equation}
\text{d}P_\star(F_\star)=rP_\star(\text{d}F_\star,F_\star, \cdots,F_\star)=rP_\star([\hat \nabla,F_\star]_s,F_\star, \cdots,F_\star)=0,
\end{equation}

\noindent where we have added a vanishing commutator term in the second equality, meanwhile use of the noncommutative Bianchi identity have been made.

\par \textbf{b)} In fact, according to the invariant theory any invariant polynomial of $K\times K$ matrices with entries in a field of characteristic zero (e.g. $\mathbb F =\mathbb R$ or $\mathbb C$) is actually composed of trace functions, hence given in terms of a polynomial of eigenvalues of matrices \cite{rasmyslev, procesi}.\footnote{It was firstly conjectured by Artin \cite{artin}.} Therefore, it is enough to prove the claim only for trace function.

\par However, as we will show below each trace polynomial could be considered as a supertrace function on some $\mathbb Z_2$-graded algebra or superalgebra. As Quillen has been shown such supertraces will provide a beautiful framework for producing many important characteristic classes within the formulation of superconnections \cite{quillen}. Hereon, we employ the spectacular machinery of Quillen's supper connection to prove the theorem.

\par Assume that $E \to X$ is a complex vector bundle, otherwise we can simply replace $E$ by $E^{\mathbb C}$. Suppose that $\nabla^t$, $t \in [0,1]$, is a continuous family of associative connections on $E$. Assume that $V$ is a complex superalgebra generated by $1$ (even) and $\{ \frak e^i \}_{i=1}^{2m}$ (odd), with $\frak e^i.\frak e^i=i \theta^{ij}$, and let $\frak T=\oplus_{k=0}^\infty V^{k \otimes}$, $ V^{0 \otimes}=\mathbb C$, be its tensor superalgebra with total degree. Now assume that $\frak S$ is another superalgebra generated by $\mathbb Z_2=\{1, \varepsilon \}$, where $\varepsilon$ is a nilpotent odd element $\varepsilon^2=1$. Hence, $S=\frak S \otimes \frak T$ is a superalgebra by the induced degree.

\par Set, $T=X\times S$ and consider $\frak E=E \otimes T$ as a super vector bundle over $X$. Put a superconnection on $\frak E$ as $\nabla_{s}=\nabla^t +{ \mathcal A}^\theta$, with
\begin{equation} \label {jaleb}
{\mathcal A}^\theta =\varepsilon F^t_i {\frak e}^i+\frac{1}{\sqrt {2!}} [\nabla^t_i,F^t_j] {\frak e}^i {\frak e}^j+ \frac{1}{\sqrt{3!}} \varepsilon [\nabla^t_i,[\nabla^t_j,F^t_k]] {\frak e}^i \frak e^j \frak e^k + \cdots,
\end{equation}

\noindent wherein $F^t_i=F^t_{i \mu} \text{d}x^\mu$ and $F^t$ is the curvature tensor of $\nabla^t$. Here we omitted the tensor product for simplicity. For example we use the short notation of $\frak e^{i_1} \frak e^{i_2} \cdots \frak e^{i_r}$ instead of $\frak e^{i_1} \otimes \frak e^{i_2} \otimes  \cdots \otimes \frak e^{i_r}$ in (\ref {jaleb}). Obviously, $\nabla_s$ is of odd degree and hence a superconnection. The curvature of $\nabla_{s}$ is given by;
\begin{equation}
\begin{split}
F_{s}=F^t-\frac{1 }{2!} \theta^{i_1 j_1} \theta^{i_2 j_2} [[\nabla^t_{i_2},F^t_{i_1}],[\nabla^t_{j_2},F^t_{j_1}]] ~~~~~~~~~~~~~~~~~~~~~~~\\
+ \frac{1}{4!} \theta^{i_1 i_1} \theta^{i_2 j_2}\theta^{i_3 j_3} \theta^{ i_4 j_4} [[\nabla^t_{i_1},[\nabla^t_{i_2},[\nabla^t_{i_3},F^t_{i_4}]]],[\nabla^t_{j_1},[\nabla^t_{j_2},[\nabla^t_{j_3}, F^t_{j_4 \nu} ]]]]+ \cdots + \mathcal O (\frak e^i; \varepsilon),   \\
\end{split}
\end{equation}

\noindent wherein $\mathcal O (\frak e^i; \varepsilon)$ includes some numbers of $\frak e_i$s or contains $\varepsilon$. Thus, upon (\ref{f-star cov});
\begin{equation}
F_{s}=F^t_{\star^t} + \mathcal O (\frak e^i; \varepsilon),
\end{equation}

\noindent for $F^t_{\star^t}$ the noncommutative curvature of $\nabla^t$ and its covariant star product $\star^t$. We obtain:
\begin{equation}
~~~~~~~~~~tr_s(F_{s}^k)=tr_s({F^t_{\star^t}}^k)=tr({F^t_{\star^t}}^k),~~~~~~~~~~k \geq 1,
\end{equation}

\noindent wherein $tr_s$ is the supertrace. Here, we used the relations $tr_s(\varepsilon)=tr_s(\frak e^i)=0$ and $tr_s(A \otimes B)=tr_s(A) tr_s(B)$. According to \cite {quillen} $tr({F^t_{\star^t}}^k)$ is a closed differential form on $X$, and its de Rham cohomology class is independent of $t$. Hence, if we put $\nabla^t=(1-t) \nabla'+t \nabla''$, for associative connections $\nabla'$ and $\nabla''$, we conclude that;
\begin{equation}
~~~~~~~~~~~~~tr(F'^k_{\star'})=tr(F''^k_{\star''})+\text{d}\lambda,~~~~~~~~~~k\geq 1,
\end{equation}

\noindent for $F'_{\star'}$ (resp. $F''_{\star''}$) the noncommutative curvature of $\nabla'$ (resp. $\nabla''$) and its covariant star product $\star'$ (resp. $\star''$). This proves the claim \textbf{(b)}.

\par \textbf{c)} To prove the claim one should employ the above machinery within some different formulation. Let $\nabla_{s}=\nabla+ t{ \mathcal A}^\theta$, $t \in [0,1]$. Following the above machinery with this superconnection leads to:
\begin{equation}
tr_s(F_s^k)=tr(F^k),~~~~~(\text{for}~t=0),~~~~~~~~~~\text{and;}
~~~~~~~~~~tr_s(F_s^k)=tr(F_\star^k),~~~~~(\text{for}~t=1),
\end{equation}

\noindent for each $k \geq 1$. Hence;
\begin{equation}
tr(F^k)=tr(F_\star^k)+\text{d}\lambda.
\end{equation}

\noindent Therefore, upon the invariant theory the de Rham cohomology classes provided by symmetric invariant polynomials for $F_\star$ coincide entirely with those obtained for $F$. This finishes the theorem.\\

\par \textbf{Another Proof by Employing the Seiberg-Witten Map;}

\noindent In this case we initially prove \textbf{(c)}, then \textbf{(b)}.

\par \textbf{c)} To establish this claim we use the elegant results of the Seiberg-Witten map. Here, we may refer to its exact formula which has significant topological features.\footnote{See \cite{banerjee1, yang} for more details.} Let $\hat \nabla^{\theta}$ be the Seiberg-Witten connection with local connection form $\hat A^\theta$ and the ordinary (commutative) curvature $\hat F^\theta$, which the later coincides with the noncommutative curvature $F_\star$ \cite{seiberg-witten}. Hence, by the Chern-Weil theorem we have: $P(F_\star)=P(\hat F^\theta)=P(F)+\text{d}\lambda$. This finishes the claim \textbf{(c)}.

\par \textbf{b)} Let $\nabla'$ be another connection on $E$. We show the covariant star product and the noncommutative curvature of $\nabla'$ respectively with $\star'$ and $F'_{\star'}$. According to \textbf{(c)} one reads;
 \begin{equation} \label {natije}
 P(F_\star)-P(F)=\text{d}\lambda,~~~~~~~~~~\text{and}~~~~~~~~~~P(F'_{\star'})-P(F')=\text{d}\lambda'.
\end{equation}

\noindent Moreover, upon the Chern-Weil theorem we have; $P(F')-P(F)=\text{d} \xi$. All in all, it is seen that:
\begin{equation}
P(F'_{\star'})-P(F_\star)=\text{d} \left(\lambda' -\lambda + \xi \right).
\end{equation}

\noindent This finishes the theorem. \textbf{Q.E.D.}


\section{Noncommutative Chern-Weil Theorem and Some Topological Information Carried by the Noncommutative Curvature}
\setcounter{equation}{0}

\par Through this and the next sections all vector bundles are assumed to be associative. Suppose that $\mathbb C^K \to E \to X$ is a complex vector bundle and $F_\star$ is its noncommutative curvature. Then, the \emph{noncommutative total Chern class} of $E$ is defined as;
\begin{equation} \label {chern-class}
c_{\star}(E)={\det} \left( 1+ \frac{i F_{\star}}{2\pi} \right):=\oplus_{k=0}^K c_{\star k}(E).
\end{equation}

\noindent Actually, $c_{\star k}(E)$, the so called \emph{$k$-th noncommutative Chern class}, is defined by;
\begin{equation} \label {n-chern-class}
c_{\star k}(E_\star)=\frac{1}{k!} \frac{\text{d}^k}{\text{d}s^k}|_{s=0} ~ {\det}\left( 1+ \frac{i s F_{\star}}{2\pi}  \right).
\end{equation}

\noindent In fact, upon to \textbf{Theorem 1} we readily find:\\

\par \textbf{Corollary 1;} \emph{For each complex vector bundle $\mathbb C^K \to E \to X$ the $k$-th noncommutative Chern class $c_{\star k}(E_\star)$, $0 \leq k \leq 2n$, is a topological invariant and cohmologous to $c_k(E)$ in $H^{2k}_{\emph{dR}}(X,\mathbb{Z})$, i.e.}
\begin{equation} \label {c-5}
\int_X c_{\star k}(E)=\int_X c_k(E) \in \mathbb{Z}.
\end{equation}

\noindent \emph{Moreover, if $E_i$s, $1\leq i \leq N$, are complex vector bundles on $X$, then:}
\begin{equation} \label {ch-class sum}
c_\star(E_1 \oplus \cdots \oplus E_N)=c_\star(E_1) \cdots   c_\star(E_N),
\end{equation}

\noindent \emph{wherein $E_1 \oplus \cdots \oplus E_N$ is the Whitney sum of vector bundles $E_i$s \cite{milnor, kobayashi}.}\\

\par A similar conclusion holds for the Chern character. In fact, if $\mathbb{F}^K \to E \to X$ is an associative vector bundle, then the \emph{$k$-th noncommutative Chern character} of $E$, $0\leq k \leq n$, is defined as:
\begin{equation} \label {n-chenr-character}
\text{ch}_{\star k}(E)=\frac{1}{k!} \text{tr} \left( \frac{i F_{\star} } {2 \pi} \right)^k .
\end{equation}

\noindent According to \textbf{Theorem 1} once again we conclude:\\

\par \textbf{Corollary 2;} \emph{For each vector bundle $\mathbb F^K \to E \to X$, $\emph{ch}_{\star k}(E)$ is a topological invariant and belongs to the integral cohomology class of $\emph{ch}_k(E)$ in $H^{2k}_{\emph{dR}}(X,\mathbb{Z})$. Specially;}
\begin{equation} \label {c-6}
\int_X \text{ch}_{\star n}(E)=\int_X \text{ch}_n(E) \in \mathbb{Z}.
\end{equation}

\noindent \emph{Moreover, if $\{E_i\}_{i=1}^N$ is a set of complex vector bundles on $X$, then:}
\begin{equation} \label {ch-class sum}
\begin{split}
\text{ch}_{\star k}(E_1 \oplus \cdots \oplus E_N)=\sum_{i=1}^N \text{ch}_{\star k}(E_i)=\text{ch}_{\star k}(E_1)+ \cdots  + \text{ch}_{\star k}(E_N),~~\\
\text{ch}_{\star k}(E_1 \otimes \cdots \otimes E_N)=\sum \prod_{i=1}^N \text{ch}_{\star k_i}(E_1) = \text{ch}_{\star k_1}(E_1) \cdots \text{ch}_{\star k_N}(E_N),
\end{split}
\end{equation}

\noindent \emph{wherein in the second equality the summation is over all $N$-plets $(k_1,\cdots,k_N)$ with $\sum_i k_i=k$.}\\

\par To study more noncommutative characteristic classes we need some more algebraic precisions. Let $\mathbb F^K \to E \to X$ be a vector bundle equipped with connection $\nabla$ so that its curvature $F$ and the noncommutative curvature $F_\star$ are both diagonalizable on $X$. For example we can assume that $E$ is a complex vector bundle and its structure group is $SU(K)$. Then, $F$ and $F_\star$ both are $\frak {su}(K)$-valued two forms and hence diagonalizable. Moreover, suppose that $\frak F(a_1,\cdots,a_K)$ is a symmetric polynomial (analytic function) of $K$ variables $a_i$s. Let $\{x_{\star i}\}_{i=1}^K$ (resp. $\{ x_{ i} \}_{i=1}^K$) be the collection of the eigenvalues of $\frac{iF_\star}{2\pi}$ (resp. $\frac{iF}{2\pi}$). Therefore, $\frak F(x_{\star 1},\cdots,x_{\star K})$ (resp. $\frak F(x_1,\cdots,x_K)$) is simply decomposed to homogeneous components by the degree of differential forms as;
\begin{equation} \label {dcmpz}
\begin{split}
\frak F(x_{\star 1},\cdots ,x_{\star K}):=\oplus_{k=0}^n \frak F_k(F_\star)=\oplus_{k=0}^n \frak F_{k}(x_{ \star 1},\cdots , x_{\star K}),\\
\frak F(x_{1},\cdots ,x_{K}):=\oplus_{k=0}^n \frak F_k(F)=\oplus_{k=0}^n \frak F_{k}(x_{ 1},\cdots , x_{ K}),~~~~~
\end{split}
\end{equation}

\noindent with $\frak F_k(a_1,\cdots,a_K)=\frac{1}{k!} \frac{\text{d}^k}{\text{d}s^k}|_{s=0}~\frak F(s a_1, \cdots , s a_K)$. Therefore, since $\frak F$ is in fact a symmetric invariant polynomial on $\mathbb M_{K\times K}(\mathbb F)$, then the noncommutative Chern-Weil theorem would be translated to the corresponding formulation based on the eigenvalues of $F_\star$.\footnote{See \cite{kobayashi} for more discussion.} For example, for the noncommutative setting we have;
$$c_\star(E)=\prod_{i=1}^K (1+x_{\star i})~~~~~~~~~~\text{and;}~~~~~~~~~~\text{ch}_\star(E)=\oplus_{k=0}^n \sum_{i=1}^K \frac{x_{\star i}^k}{k!}=\sum_{i=1}^K e^{x_{\star i}}.$$
\noindent The following corollary provides a noncommutative formulation of Chern-Weil theorem given in terms of eigenvalues of $F_\star$.\\

\par \textbf{Corollary 3;} \emph{For each vector bundle $\mathbb F^K \to E \to X$, and for any symmetric polynomial $\frak F(a_1,\cdots,a_K)$, the homogeneous polynomials $\frak F_k(F)$ and $\frak F_k(F_{\star})$, $0\leq k \leq n$, defined by (\ref {dcmpz}), are closed and belong to the same de Rham cohomology class of $H^{2k}_{\emph{dR}}(X,\mathbb R)$. In fact, $\frak F_k(F_\star)$ is a topological invariant and its cohomology class is independent of the connection of $E$.}\\

\par Now, by employing \textbf{Corollary 3} we can provide the noncommutative versions of more characteristic classes. For instance, the \emph{noncommutative total Todd class} is defined as;
\begin{equation} \label {NC-todd-class}
\text{Td}_{\star}(E)=\frac{1}{K!} \sum_{\sigma \in S_K} \left( \frac{x_{\star  \sigma(1)}}{1-e^{-x_{\star \sigma(1)}}} \right) \cdots \left( \frac{x_{\star \sigma(K)}}{1-e^{-x_{\star \sigma(K)}}} \right),
\end{equation}

\noindent wherein $S_K$ is the permutation group of $\{1,\cdots,K\}$. In principle, the noncommutative Todd class is decomposed as; $\text{Td}_{\star}(E)=\oplus_{k=0}^n \text{Td}_{\star k}(E)$, wherein $\text{Td}_{\star k}(E)$ is referred to as the \emph{$k$-th noncommutative Todd class} of $E$.\\ 

\par \textbf{Corollary 4;} \emph{For each vector bundle $\mathbb F^K \to E \to X$ the $k$-th noncommutative Todd class of $E$, $\text{Td}_{\star k}(E)$, $0 \leq k \leq n$, is a topological invariant and belongs to the rational cohomology class of $\text{Td}_k(E)$ in $H^{2k}_{\text{dR}}(X,\mathbb Q)$ \cite{hirz-book}. On the other hand, if $\{E_i\}_{i=1}^N$ is a set of vector bundles over $X$, then;}
\begin{equation}
\text{Td}_{\star}(E_1 \oplus \cdots \oplus E_N)=\prod_{i=1}^N \text{Td}_{\star}(E_i)=\text{Td}_{\star}(E_1) \cdots \text{Td}_{\star}(E_N).
\end{equation}

\noindent \emph{Moreover, if $X$ is a complex manifold, then the arithmetic genus of $X$ is given as;}
\begin{equation} \label {corollary 7-1}
\sum_{k=0}^n (-1)^k b^{0,k}=\int_X \text{Td}_{\star n} (TX^+),
\end{equation}

\noindent \emph{wherein $b^{0,k}=\dim_{\mathbb C}H^{0,k}(X,\mathbb C)$, $0\leq k \leq n$, are the Hodge numbers, and $TX^+$ is the holomorphic tangent bundle of $X$.}\\

\par Similarly, the \emph{noncommutative Hirzebruch $L$-polynomial} is defined as;
\begin{equation} \label {NC-hirzebruch}
L_{\star}(E)=\frac{1}{K!} \sum_{\sigma \in S_K} \left( \frac{x_{\star  \sigma(1)}}{\tanh({x_{\star \sigma(1)})}} \right)  \cdots \left( \frac{x_{\star \sigma(K)}}{\tanh({x_{\star \sigma(K)})}} \right):= \oplus_{k=0}^{[n/2]} L_{\star k}(E).
\end{equation}

\noindent As a convention $L_{\star k}(E)$ is referred to as the \emph{$k$-th noncommutative Hirzebruch $L$-genus}, while the highest term of (\ref {NC-hirzebruch}), is called the \emph{top noncommutative Hirzebruch $L$-genus}. It can be seen that $L_{\star k}(E)$ is a $4k$-form with $0 \leq k \leq n/2$. Hence, to obtain a volume form from the Hirzebruch polynomial $n$ must be an even number, say $n=2p$.  \\

\par \textbf{Corollary 5;} \emph{For each vector bundle $\mathbb F^K \to E \to X$, the $k$-th noncommutative Hirzebruch genus $L_{\star k}(E)$, $0 \leq k \leq n/2$, is a topological invariant and cohomologous to $L_k(E)$ in $H_{dR}^{4k}(X,\mathbb Z)$. Specially, if $n=2p$, then the top noncommutative Hirzebruch $L$-genus $L_{\star p}(E)$ is a topological invariant and}
\begin{equation} \label {corollary 8}
\int_X L_{\star p}(E)=\int_X L_p(E) \in \mathbb Z.
\end{equation}

\noindent \emph{On the other hand, if $X$ is a 4-manifold and $E=TX$, then;}
\begin{equation} \label {corollary 8-1}
\tau(X)=\int_X L_{\star 1}(TX) \in \mathbb Z,
\end{equation}
\noindent \emph{wherein $\tau(X)$ is the Hirzebruch signature. Moreover, if $X$ is also a spin manifold, then by Rochlin's theorem \cite {rokhlin} the noncommutative $L$ genus $L_{\star 1}(TX)$ (i.e. the result of (\ref {corollary 8-1})) is divided by 16 \cite{freedman}.}\\

\par We emphasize that upon \textbf{Lemma 1} the Hirzebruch signature of $X$ as a $2n$-dimensional manifold could be calculated in the $\star$ formulation of pairing $\left\langle {.}
 \mathrel{\left | {\vphantom { . .}} \right. \kern-\nulldelimiterspace} {.} \right\rangle_\star:H^n_{\text{dR}}(X,\mathbb R) \otimes H^n_{\text{dR}}(X,\mathbb R) \to \mathbb R$, as;
\begin{equation} \label {hirzebruch-star-pairing}
\left[ \omega_1 \right]
 \otimes [ \omega_2 ] \mapsto
 \left\langle {\omega_1}
 \mathrel{\left | {\vphantom { \omega_1 \omega_2}}
 \right. \kern-\nulldelimiterspace}
 {\omega_2} \right\rangle_\star:=\int_X \omega_1 \star \omega_2=\int_X \omega_1 \wedge \omega_2= \left\langle {\omega_1}
 \mathrel{\left | {\vphantom { \omega_1 \omega_2}}
 \right. \kern-\nulldelimiterspace}
 {\omega_2} \right\rangle.
\end{equation}

\noindent In this sense the noncommutative Hirzebruch signature $\tau_\star(X)$ coincides with the ordinary Hirzebruch signature $\tau(X)$, and hence, (\ref {corollary 8-1}) could be interpreted as a significant evidence of a noncommutative index theorem on noncommutative manifolds.

\par The \emph{noncommutative Dirac genus} (or the \emph{noncommutative $\hat A$ genus}) is also defined with;
\begin{equation} \label {NC-A}
\hat A_{\star}(E)=\oplus_{k=0}^{[n/2] }\hat A_{\star k}(E)=\frac{1}{K!} \sum_{\sigma \in S_K} \left( \frac{x_{\star \sigma(1)}/2}{\sinh({x_{\star \sigma(1)}/2 )}} \right)  \cdots  \left( \frac{x_{\star \sigma(K)}/2}{\sinh({x_{\star \sigma(K)}/2) }} \right).
\end{equation}

\noindent Each $\hat A_{\star k}(E)$ is referred to as the \emph{$k$-th noncommutative Dirac genus}. Similar to the Hirzebruch polynomial, it can be seen that the noncommutative Dirac genus also consists of $4k$-forms for $0\leq k \leq n/2$, i.e. $\hat A_{\star k}(E)$. Hence, once again, to obtain a topological invariant volume element, one has to assume that $n=2p$ for some integer $p$. Then, $\hat A_{\star p}$, the \emph{top noncommutative Dirac genus} exists and could be integrated over $X$.\\

\par \textbf{Corollary 6;} \emph{For each vector bundle $\mathbb F^K \to E \to X$, the $k$-th noncommutative Dirac genus $\hat A_{\star k}(E)$  is a topological invariant and belongs to the integral cohomology class of $\hat A_k(E)$ in $H^{4k}_{\text{dR}}(X,\mathbb Q)$. Specially, if $n=2p$, then the top noncommutative Dirac genus of $E$ is a topological invariant and}
\begin{equation} \label {corollary 9}
\int_X \hat A_{\star p}(E)=\int_X \hat A_p(E) \in \mathbb Q.
\end{equation}

\noindent \emph{Moreover, according to the Borel-Hirzebruch theorem \cite{bh, bh2} if $X$ is a spin manifold $A_{\star k}$ belongs to integral cohomology group $H^{4k}_{\text{dR}}(X,\mathbb Z)$, and if $p$ is odd;}
\begin{equation} \label {corollary 9-1}
\int_X \hat A_{\star p}(TX) \in 2 \mathbb Z,
\end{equation}

\noindent \emph{while if the above result is non-zero, then due to the Lichnerowicz obstruction theorem \cite{lic}, $X$ would not admit a metric $g_{\mu \nu}$ with positive scalar curvature.}\\

\par If in addition the vector bundle $\mathbb R^{K} \to E \to X$ is orientable and endowed with a smooth metric on the fibers, say $\frak g_{pq}$, and a metric compatible connection $\nabla$ (i.e. $\nabla \frak g_{pq}=0$), then the \emph{noncommutative total Pontrjagin class} is defined by;
\begin{equation} \label {NC-pontrjagin}
p_{\star}(E)={\det}_{\star} \left(1+\frac{ F_{\star} }{2\pi} \right):=\oplus_{k=0}^{n} p_{\star k/2}(E),
\end{equation}

\noindent wherein $p_{\star k/2}(E_\star)$, the so called \emph{$\frac{k}2$-th noncommutative Pontrjagin class} is a $2k$-form. Also, similar to the Chern class, the noncommutative Pontrjagin class would not provide a volume element unless $K \geq n$. Hence, we may simply assume that $K \geq n$. One should note that, just similar to the commutative case, the half integer noncommutative Pontrjagin classes vanish identically; $p_{\star k/2}(E)=p_{k/2}(E)=0$ for $k=1 $ mod $2$. Specially, the \emph{top noncommutative Pontrjagin class} $p_{\star n/2}(E_\star)$ vanishes for odd $n$. In fact, for real vector bundle $\mathbb R^{K} \to E\to X$ and the structure group $SO(K)$, both $F$ and $F_\star$, are $\frak {so}(K)$-valued two forms, hence traceless. For instance if $K=2k$ is even, then we can block-diagonalize $\frac F {2\pi}$ and $\frac {F_\star} {2\pi}$ as:
\begin{equation} \label {semi-diag}
F=\left[ {\begin{array}{*{20}{c}}
   {\begin{array}{*{20}{c}}
   0 & x_1  \\
   { - x_1} & 0  \\
\end{array}} & {} & {}  \\
   {} &  \ddots  & {}  \\
  {} & {} & {\begin{array}{*{20}{c}}
   0 & x_k  \\
   { - x_k} & 0  \\
\end{array}}  \\
\end{array}} \right]
~~~~~~~~~~\text{and;}~~~~~~~~~~F_\star=\left[ {\begin{array}{*{20}{c}}
   {\begin{array}{*{20}{c}}
   0 & x_{\star 1}  \\
   { - x_{\star 1}} & 0  \\
\end{array}} & {} & {}  \\
   {} &  \ddots  & {}  \\
   {} & {} & {\begin{array}{*{20}{c}}
   0 & x_{\star k}  \\
   { - x_{\star k}} & 0  \\
\end{array}}  \\
\end{array}} \right].
\end{equation} 

\par In fact, the generating function of $p_\star(E)$ is given by;
\begin{equation} \label {generating pontrjagin}
p_\star(E)=\prod_{i=1}^{[K/2]} (1+x_{\star i}^2),~~~~~~~~~~\text{with;}~~~~~~~~~~p_{\star k}=\sum x_{\star i_1}^2 x_{\star i_2}^2 \cdots x_{\star i_k}^2,
\end{equation}

\noindent wherein the summation in the second equality is over all ${1\leq i_1<i_2<\cdots <i_k \leq [K/2]}$ and $x_{\star i}$s are two-form eigenvalues of $\frac {F_\star}{2 \pi}$ introduced in (\ref {semi-diag}). \\

\par \textbf{Corollary 7;} \emph{For each Riemannian real vector bundle $\mathbb R^K \to E \to X$, the $\frac {k}2$-th noncommutative Pontrjagin class $p_{\star k/2}(E)$ is a topological invariant and is cohomologous to $p_{k/2}(E)$ in the integral cohomology group of $H^{2k}_{\text{dR}}(X,\mathbb Z)$. Specially, if $n=2p$ and $K\geq n$, then the top noncommutative Pontrjagin class $p_{\star p}(E)$ is cohomologous to $p_{p}(E)$ in $H^{2n}_{\text{dR}}(X,\mathbb Z)$ and}
\begin{equation} \label {corollary 10}
\int_X p_{\star n/2}(E_\star)=\int_X p_{n/2}(E) \in \mathbb Z.
\end{equation}

\noindent \emph{For this situation $p_{\star p}(E)$, $(-1)^p c_{\star n}(E^{\mathbb C})$ and $(-1)^p c_n(E^{\mathbb C})$ all are cohomologous, wherein $E^{\mathbb C}$ is the complexification of $E$ \cite{milnor}. Moreover, if $K=2l$ or $2l+1$, then for each set of integers $\{ a_i \}_{i=1}^{l}$ with weighted sum $a_1+2a_2+\cdots + k a_k =p$, the integral of}
\begin{equation} \label {ponrjagin number}
\int_X p_{\star 1}^{a_1} (E) p_{\star 2}^{a_2} (E) \cdots  p_{\star l}^{a_l}(E)
\end{equation}

\noindent \emph{coincides with the Pontrjagin number of $E$ for $l$-plet $(a_1,\cdots,a_l)$ \cite{tu}.}\\

\par The \emph{noncommutative Euler class} is similarly defined for the real vector bundle $\mathbb R^K \to E \to X$ as;
\begin{equation} \label {NC-euler-class}
e_{\star}(E)=\text{Pf} \left(\frac{F_{\star}}{2\pi} \right)=\frac{(-1)^K}{2^K K!} \epsilon^{a_1 \cdots a_{2K}} \left(\frac{F_{\star}}{2\pi} \right)_{a_1 a_2} \cdots \left(\frac{F_{\star}}{2\pi} \right)_{a_{2K-1} a_{2K}},
\end{equation} 

\noindent wherein $a_i$s are the indices of $\frac{F_{\star}}{2\pi} \in \mathbb M_{K \times K}(\mathbb R)$. Obviously, $e_{\star'} (E_\star)$ would be a volume element only for $n=K$. For instance $e_\star(TX)=x_{\star 1} \cdots x_{\star n}$. \\

\par \textbf{Corollary 8;} \emph{Assume that $\mathbb R^K \to E \to X$ is a real vector bundle. Then, the noncommutative Euler class $e_{\star}(E)$ is a topological invariant and belongs to the integral cohomology class of $e(E)$ in $H^{2K}_{\emph{dR}}(X,\mathbb{Z})$. Specially, if $K=n$, then;}
\begin{equation} \label {corollary 11}
\int_X e_{\star}(E)=\int_X e(E) \in \mathbb{Z}.
\end{equation}


\section{Associative Vector Bundles, Noncommutative Elliptic Operators and Noncommutative Topological Index Theorem}
\setcounter{equation}{0}

\par Following the statements of the last section leads us naturally to obtain a noncommutative version of the celebrated Atiyah-Singer index theorem.\footnote{As mentioned before, through this section all vector bundles are assumed to be associative.} For instance, the \textbf{Corollary 8} naturally provides the noncommutative Gauss-Bonnet theorem:\\

\par \textbf{Corollary 9;} \emph{Assume that $e_{\star}(TX)$ is the noncommutative Euler class of the tangent bundle $TX$. Then;}
\begin{equation} \label {corollary 11-1}
\int_X e_{\star} (TX )= \chi(X) \in \mathbb Z,
\end{equation}

\noindent \emph{wherein $\chi (X)$ is the Euler characteristic of $X$. Moreover, if $\mathcal X$ is a smooth vector field on $X$ with finitely many zeros and index sum $j(\mathcal X)$, then \cite{curv-conn};}
\begin{equation} \label {index-field}
j(\mathcal X)=\int_X e_{\star}(TX_\star).
\end{equation}
~
\par Actually, since the star product $\star$ never touches the CW-complex structure of $X$, or the differential setting of the de Rham cohomology classes,\footnote{It appears only as a deformation of the cup product of the de Rham cohomology classes.} we see that the noncommutative Euler character $\chi_\star(X)$, obtained by pairing $e_\star(TX)$ and the fundamental homology class of $X$, coincides with the ordinary Euler character $\chi(X)$ via;
$$\chi_\star(X) =\chi(X)= \sum_{k=0}^{2n} (-1)^k \#(\text{\emph{k}-dimensional cells})=\sum_{k=0}^{2n} (-1)^k \dim H^k_{\text{dR}}(X,\mathbb R),$$
\noindent hence, \textbf{Corollary 9} could be understood as important evidence for existing a pure formulation of noncommutative topological index theorem.\footnote{Actually, what we are looking for is a topological noncommutative index theorem which, of course, needs its own construction independent of the algebraic noncommutative index theorem introduced in noncommutative geometry \cite {connes}. See the viewpoints of \cite{singer, tellman} for more discussions.} Moreover, as we stated above, the noncommutative Hirzebruch signature theorem, reported in \textbf{Corollary 5}, provides another clue for such an existence. To work out a noncommutative topological index theorem we need some more preliminaries.

\par Assume that $\mathbb C^K\to E \to X$ is a complex vector bundle endowed with a Hermitian inner product. This inner product will induce an inner product on the vector space of smooth sections of $E$, i.e. $\Gamma(E)$. Let us show it with $\left\langle {.} \mathrel{\left | {\vphantom {. .}} \right. \kern-\nulldelimiterspace} {.} \right\rangle$;
\begin{equation} \label {C-inner product}
~~~~~~~~~~~~~~~~~~~~\left\langle {\zeta} \mathrel{\left | {\vphantom {. .}} \right. \kern-\nulldelimiterspace} {\zeta'} \right\rangle= \int_X * \left(\overline {\zeta_p} \zeta'^p \right),~~~~~~~~~~\zeta, \zeta' \in \Gamma(E).
\end{equation}

\par This, will put a noncommutative structure on $E$, say $\left\langle {.} \mathrel{\left | {\vphantom {. .}} \right. \kern-\nulldelimiterspace} {.} \right\rangle_{\star}$, by employing (\ref {inner-associative});
\begin{equation} \label {NC-inner product}
\left\langle {\zeta} \mathrel{\left | {\vphantom {. .}} \right. \kern-\nulldelimiterspace} {\zeta'} \right\rangle_{\star}:= \int_X *\left( \overline {\zeta_p} \star \zeta'^p \right)=\int_X * \left(\overline {\zeta_p} \zeta'^p \right)=\left\langle {\zeta} \mathrel{\left | {\vphantom {. .}} \right. \kern-\nulldelimiterspace} {\zeta'} \right\rangle,
\end{equation}

\noindent hence, $\left\langle {.} \mathrel{\left | {\vphantom {. .}} \right. \kern-\nulldelimiterspace} {.} \right\rangle_{\star} = \left\langle {.} \mathrel{\left | {\vphantom {. .}} \right. \kern-\nulldelimiterspace} {.} \right\rangle$. Thus, the ordinary and the noncommutative norms coincide, i.e.;
\begin{equation} \label {norm====norm}
~~~~~~~~~~|\zeta|_{\star} =\sqrt{\left\langle {\zeta} \mathrel{\left | {\vphantom {. .}} \right. \kern-\nulldelimiterspace} {\zeta} \right\rangle_{\star}}=\sqrt{\left\langle {\zeta} \mathrel{\left | {\vphantom {. .}} \right. \kern-\nulldelimiterspace} {\zeta} \right\rangle}= |\zeta|,~~~~~~~~~~\zeta \in \Gamma(E).
\end{equation}

\par Now, suppose that $\mathbb C^{K} \to E \to X$ and $\mathbb C^{K'} \to E' \to X$ are two complex vector bundles over $X$ and $D:\Gamma(E) \to \Gamma(E')$ be a linear pseudo-differential operator with noncommutative version of $D_\star$, defined with (\ref {D-3}). The noncommutative norm of $D$ is defined as the norm of $D_\star$ i.e.;
\begin{equation} \label{NC-norm}
|D_\star|=\text{Sup}_{\zeta \in S(E)} |D_\star\zeta|=\text{Sup}_{\zeta \in S(E)} |D_\star \zeta|,
\end{equation}

\noindent wherein $S(E)=\{ \zeta \in \Gamma(E) | |\zeta|=1 \}$ is the unit sphere of $\Gamma(E)$. The following lemma shows that $D_\star$ inherits the most significant properties of $D$. \\

\par \textbf{Lemma 4;} \emph{The linear pseudo-Differential operator $D_\star:\Gamma(E) \to \Gamma(E')$ is a bounded linear map, if and only if $D$ is bounded too. Moreover, $|D_\star \zeta|=|D\zeta|$ for each $\zeta \in \Gamma(E)$, hence; $|D_\star|=|D|$.}\\

\par \textbf{Proof;} Initially, we note that upon to (\ref {NC-inner product}) the noncommutative setting is furnished by the Cauchy-Schwarz inequality;
\begin{equation} \label {schwarz}
|\left\langle {\zeta} \mathrel{\left | {\vphantom {. .}} \right. \kern-\nulldelimiterspace} {\zeta'} \right\rangle_{\star}| \leq |\zeta| |\zeta'|.
\end{equation}

\noindent Let $\zeta \in \Gamma(E)$ be an arbitrary smooth section. We have;
\begin{equation} \label {d star , d}
|D_\star \zeta|^2=\sqrt{\left\langle {D_\star\zeta} \mathrel{\left | {\vphantom {. .}} \right. \kern-\nulldelimiterspace} {D_\star \zeta} \right\rangle}=\sqrt{\left\langle {D_\star\zeta} \mathrel{\left | {\vphantom {. .}} \right. \kern-\nulldelimiterspace} {D \zeta} \right\rangle}_\star \leq |D_\star \zeta| |D \zeta|,
\end{equation}

\noindent wherein in the second equality use \textbf{Lemma 1} has been made. Hence, we obtain;
\begin{equation} \label {ddd'}
|D_\star \zeta| \leq |D \zeta |.
\end{equation}

\noindent Obviously, (\ref {ddd'}) could be proved for $\star^{-1}$ which is defined by $-\theta^{ij}$. Hence, if we replace $D$ and $\star$ respectively by $D_\star$ and $\star^{-1}$, we would find:
\begin{equation} \label {ddd''}
|D \zeta | \leq |D_\star \zeta|.
\end{equation}

\noindent Consequently, upon (\ref {ddd'}) and (\ref {ddd''}) for each $\zeta \in \Gamma(E)$ we have:
\begin{equation} \label {karamad}
|D_\star \zeta|=|D \zeta|,
\end{equation}

\noindent which yields: $|D_\star|=|D|$. This finishes the lemma. \textbf{Q.E.D.}\\

\par Now, we are ready to work out our main objective. The following theorem is the noncommutative version of the celebrated Atiyah-Singer index theorem.  \\

\par \textbf{Theorem 2;} \emph{Let $\mathbb F^K \to E \to X$ and $\mathbb F^K \to E' \to X$, be two smooth vector bundles over $X$ and $D: \Gamma(E) \to \Gamma(E')$ be an elliptic pseudo-differential operator. Then, $D_\star$ is also elliptic and;}
\begin{equation} \label {theorem2}
\text{Index}(D_\star)=(-1)^{n} \int_X \left( \text{ch}_{\star}( E) -\text{ch}_{\star}(E)  \right) \left( \frac{ \text{Td}_{\star} ( TX^{\mathbb C} ) }{e_{\star} (TX)} \right) \mid_{\text{vol}},
\end{equation}

\noindent \emph{wherein;}
$$\frac{ \text{Td}_\star ( TX^{\mathbb C} ) }{e_\star (TX)}=\oplus_{k=0}^n \left( \frac{\text{Td}_\star (TX^{\mathbb C}  )    }{e_\star (TX)} \right)_k,$$
\noindent \emph{is the noncommutative total Todd class modulo the noncommutative Euler class, as is defined in the ordinary differential geometry \cite{atiyah68-2}.}\\

\par \textbf{Proof;} First of all we have to prove the following two statements:

\par \textbf{a)} \emph{If $D$ is elliptic, then $D_\star:\Gamma(E_\star) \to \Gamma(E'_\star)$ is also an elliptic operator.}
\par \textbf{b)} \emph{If $D:\Gamma(E) \to \Gamma(E')$ is elliptic, then we have;}
\begin{equation} \label {lemma 7}
\text{Index}(D_\star)=\text{Index}(D).
\end{equation}

\noindent We prove the above claims simultaneously. According to \textbf{Lemma 4} we find:
\begin{equation} \label {desdes}
\begin{split}
Ker(D_\star)=\{ \zeta \in \Gamma (E)| D_\star \zeta=0 \}
=\{ \zeta \in \Gamma(E)| |D_\star \zeta |=0\}\\
=\{\zeta \in \Gamma(E)| |D \zeta|=0\}
=\{ \zeta \in \Gamma(E)| D \zeta=0 \}=Ker(D).
\end{split}
\end{equation}

\noindent On the other hand, due to (\ref {inner-associative}) we readily obtain: ${ D_\star }^\dag=\left(D^\dag\right)_\star$. Therefore, similar to the reasoning of (\ref {desdes}) we can prove:
\begin{equation} \label {des}
Ker( {D_\star}^\dag )=Ker( D^\dag).
\end{equation}

\noindent Thus;
\begin{equation} \label {fred-1}
\dim Ker(D_\star)=\dim Ker(D) < \infty,~~~~~~~\dim Ker({D_\star}^\dag)=\dim Ker (D^\dag) <\infty.
\end{equation}

\par Now we establish that $Im D_\star$ is closed. Assume that $\{ \xi_i \}_{i=1}^\infty$ is a convergent sequence in $Im D_\star$. Therefore, there exists a sequence in $\Gamma(E)$, say $\{ \zeta_i \}_{i=1}^\infty$, so that $D_\star \zeta_i=\xi_i$. On the other hand, since $|D(\zeta_m-\zeta_n)|=|D_\star(\zeta_m-\zeta_n)|$, then the image of $\{ \zeta_i \}$ under the action of $D$ is a Cauchy sequence, and convergent to some point in $Im D$, say $D \zeta$. We claim that $\xi_i \to D_\star \zeta$, hence $Im D_\star$ is closed. In fact, because of the identity
$$|D_\star(\zeta - \zeta_i)|=|D(\zeta -\zeta_i)|$$
\noindent the Cauchy sequence $D_\star \zeta_i$ must converge to $D_\star \zeta$, which proves the closedness of $Im D_\star$. Actually, the same reasoning holds for ${D_\star}^\dag$, hence $Im D_\star$ is closed too. Thus, according to (\ref {fred-1}) and the closedness of $Im D_\star$ and $Im {D_\star}^\dag$ we conclude that $D_\star \in \text{Hom}(\Gamma(E),\Gamma(E'))$ is a Fredholm operator, hence elliptic \cite{gilkey}. Moreover, from (\ref {desdes}) and (\ref {des}) we read:
$$\text{Index}(D_\star)=\text{Index}(D),$$
\noindent which proves \textbf{(a)} and \textbf{(b)}.

\par In principle, one can demonstrate the identity (\ref {lemma 7}) by considering the homotopy invariance of index of elliptic operators. Through this way we do not refer to (\ref {desdes}) and (\ref {des}), but instead we only insist on the ellipticity of $D_\star \in \text{Hom}(\Gamma(E),\Gamma(E'))$. Let us define a homotopy of operators in $\text{Hom}(\Gamma(E),\Gamma(E'))$ with replacing $\theta^{ij} \to s \theta^{ij}$ and $\star \to \star_s$ for $s\in [0,1]$. Then, since the ellipticity does not depend on the values of $\theta^{ij}$ we conclude that $D_{\star_s} \in \text{Fred}(\Gamma(E),\Gamma(E'))$ for all $s \in [0.1]$. Therefore, $\gamma(s):=D_{\star_s}:[0,1] \to \text{Fred}(\Gamma(E_\star),\Gamma(E'_\star))$ is a smooth curve which initiates at $D$ and terminates in $D_\star$. Since, the index of Fredholm operators is invariant via homotopic relations \cite{gilkey}, then (\ref {lemma 7}) is obtained.

\par Now we are ready to prove (\ref {theorem2}). Actually, according to (\ref {lemma 7}) and \textbf{Theorem 1}, and upon the Atiyah-Singer index theorem \cite{atiyah1, atiyah68-2} we readily obtain the following equalities;
\begin{equation} \label {th2-proof}
\begin{gathered}
\text{Index}(D_\star)=\text{Index}(D) ~~~~~~~~~~~~~~~~~~~~~~~~~~~~~~~~~~~~~\\
~~~~~~~~~~~~~~~~~~~~~~~~~=(-1)^{n} \int_X \left( \text{ch}( E) -\text{ch}(E')  \right) \left( \frac{ \text{Td} ( TX^{\mathbb C} ) }{e (TX)} \right) \mid_{\text{vol}} ~~\\
~~~~~~~~~~~~~~~~~~~~~~~~~~~~~~~~~~=(-1)^{n} \int_X \left( \text{ch}_{\star}( E) -\text{ch}_{\star}(E')  \right) \left( \frac{ \text{Td}_{\star} ( TX^{\mathbb C} ) }{e_{\star}(TX)} \right) \mid_{\text{vol}}.~~~~~~~
\end{gathered}
\end{equation}

\noindent Hence, the theorem follows. \textbf{Q.E.D.}\\

\par Actually, the noncommutative topological index theorem must be generalized to elliptic complexes on $X$. In principle, if
\begin{equation} \label {ell-com}
\cdots \xrightarrow[\text{~}]{D_{i-2}}  \Gamma(E_{i-1})  \xrightarrow[\text{~}]{D_{i-1}}  \Gamma(E_{i})  \xrightarrow[\text{~}]{D_{i}} \Gamma(E_{i+1})  \xrightarrow[\text{~}]{D_{i+1}} \cdots 
\end{equation}

\noindent is an elliptic complex, then its noncommutative version is defined as; 
\begin{equation} \label {NC-ell-com}
\cdots \xrightarrow[\text{~}]{D_{i-2 \star}}  \Gamma(E_{i-1})  \xrightarrow[\text{~}]{D_{i-1 \star}}  \Gamma(E_{i })  \xrightarrow[\text{~}]{D_{i \star}} \Gamma(E_{i+1})  \xrightarrow[\text{~}]{D_{i+1 \star}} \cdots .
\end{equation}

\noindent Actually, according to \textbf{Lemma 4} and the proof of \textbf{Theorem 1} we can see that; $Im D=Im D_\star$. Thus, for each $\zeta \in \Gamma(E_i)$ there exists some $\zeta' \in \Gamma(E_i)$ so that; $D_i \zeta' =D_{i \star} \zeta$. Moreover, upon \textbf{Lemma 4} we know that $\xi \in Ker D_{i+1}$ if and only if it belongs to $Ker D_{i+1 \star}$. All in all, we conclude:
\begin{equation} \label {exactness}
\left( D_{i+1 \star} \circ D_{i \star} \right)\zeta= D_{i+1 \star}(D_{i }\zeta')=0,
\end{equation}

\noindent for each $\zeta \in \Gamma(E_i)$. Therefore, if (\ref {ell-com}) is an elliptic complex, then (\ref {NC-ell-com}) is elliptic too. Altogether, we obtain:
\begin{equation} \label {cohom}
H^i_\star(E,D)=Ker D_{i \star}/Im D_{i-1 \star}=Ker D_i/Im D_{i-1}=H^i(E,D)
\end{equation}

\noindent Consequently;
\begin{equation} \label {in-ell-com}
\text{Index}(E_i,D_{i \star})=\sum_i (-1)^i \dim H^i_\star(E,D)=\sum_i (-1)^i \dim H^i(E,D)=\text{Index}(E_i,D_i),
\end{equation}

\noindent which is an equivalent formulation of the noncommutative index theorem for elliptic complexes. For instance, the noncommutative version of the de Rham complex 
\begin{equation} \label {derham-ell-com}
\cdots \xrightarrow[\text{~}]{d_{i-2 \star}}  \Omega^{i-1}(X)  \xrightarrow[\text{~}]{d_{i-1 \star}}  \Omega^i(X)  \xrightarrow[\text{~}]{d_{i \star}} \Omega^{i+1}(X)  \xrightarrow[\text{~}]{d_{i+1 \star}} \cdots 
\end{equation}

\noindent fulfills the nilpotency condition $d_{i \star} \star d_{i-1 \star} =0$, and $H^i_{dR \star}(X,\mathbb F)=H^i_{dR}(X,\mathbb F)$. Thus, according to the above statements we read:
\begin{equation}
\text{Index}(\Omega^i(X),\text{d}_{i \star})=\sum_i (-1)^i \dim H^i_{dR \star}(X, \mathbb F)=\sum_i (-1)^i \dim H^i_{dR}(X,\mathbb F)=\text{Index}(\Omega^i(X),\text{d}_i),
\end{equation}

\noindent which coincides with $\chi (X)$ in agreement with \textbf{Corollary 9}. However, with employing the rolling up trick to sum the even and the odd bundles, once again we are encountered with an elliptic operator $D$ and its noncommutative version $D_\star$, i.e.;
\begin{equation} \label {NC-rolling up}
0 \hookrightarrow  \Gamma(E^{\text{even}}_\star)  \xrightarrow[\text{~}]{D_\star}  \Gamma(E^{\text{odd}}_\star)  \xrightarrow[\text{~}]{0} 0,
\end{equation}

\noindent wherein $E^{\text{even}}=\oplus_i E_{2i}$, $E^{\text{odd}}=\oplus_i E_{2i+1}$, and $D=\oplus_i \left( D_{2i} + D^\dag_{2i-1}  \right)$. For this rolling up bundle we can always use the noncommutative topological index theorem via the following equality;
\begin{equation}
\text{Index}(D_\star)=\text{Index}(D)=\text{Index}(E_i,D_i).
\end{equation}

\par Hence, we have the following corollary.\\

\par \textbf{Corollary 10;} \emph{The index of the noncommutative version of the de Rham operator $\text{d}_X$, considered for the rolled up vector bundles $\wedge^{\text{even}}T^*X$ and $\wedge^{\text{odd}}T^*X$ is given by;}
\begin{equation} \label {corollary 16}
\begin{gathered}
\text{Index}(\Omega^i(X),\text{d}_{i\star})\\
=(-1)^{n} \int_X  \left( \text{ch}_{\star} ( \wedge^{\text{even}}T^*X) -\text{ch}_{\star}(\wedge^{\text{odd}} T^*X)  \right) \left( \frac{ \text{Td}_{\star} ( TX^{\mathbb C} ) }{e_{\star} (TX)} \right) \mid_{\text{vol}}=\int_X e_{\star} (TX).
\end{gathered}
\end{equation}

\noindent \emph{In addition, according to \textbf{Corollary 9} this index coincides with the Euler character of $X$, i.e.;}
\begin{equation} \label {c16-1}
\text{Index}(\text{d}_{ \star})=\int_X e_{\star} (TX) =\chi(X)=\sum_k (-1)^k b^k,
\end{equation}

\noindent \emph{wherein $b^k=\dim H^k_{\text{dR}}(X,\mathbb R)$, $0 \leq k \leq 2n$, are the Betti numbers.}\\ 

\par Now let us have a short review on the preliminaries of the Hirzebruch signature theorem and fix the notations due. Let $\Omega^*(X)=\oplus_{k=0}^{2n} \Omega^k(X)$, be the graded algebra of differential forms on $X$, and define $\eta:\Omega^*(X) \to \Omega^*(X)$ with $\eta \mid_{  \Omega^k(X)}: \Omega^k (X) \to \Omega^{2n-k}(X)$, as; $\eta=i^{k(k-1)+n} ~\textbf{*}$, which fulfills $\eta^2=1$ and $[D,\eta]=0$, for $D=\text{d}+\text{d}^\dag$. The operator $\eta$ has two eigenvalues $\pm 1$, which the corresponding eigenspaces are respectively denoted by $\Omega^+(X)$ and $\Omega^-(X)$. In fact, $D^+$, the restriction of $D$ to $\Omega^+(X)$, is an elliptic operator from the rolled up bundle $\Omega^+(X)$ to $\Omega^-(X)$. Thus, the noncommutative operator $D^+_{\star}:\Omega^+(X) \to \Omega^-(X)$ is an elliptic operator too and $\text{Index}(D^+_\star)=\text{Index}(D^+)$. The next corollary is the noncommutative version of the Hirzebruch signature theorem.
\\

\par \textbf{Corollary 11;} \emph{The index of $D^+_\star$ is given with the following formula;}
\begin{equation} \label {hc1-1}
\begin{gathered}
\text{Index}(D^+_\star)\\
=
 (-1)^{n}\int_X  \left( \text{ch}_{\star}( \wedge^+T^*X) -\text{ch}_{\star}(\wedge^- T^*X)  \right) \left( \frac{ \text{Td}_{\star} ( TX^{\mathbb C} ) }{e_{\star}(TX)} \right) \mid_{\text{vol}}=\int_X L_{\star} (TX)\mid_{\text{vol}}.
 \end{gathered}
\end{equation}

\noindent \emph{More precisely, according to \textbf{Corollary 5} this index is equal to the Hirzebruch signature of $X$, i.e.;}
\begin{equation} \label {hc1-2}
\tau(X)=b_+-b_-=\text{Index}(D^+_\star)=\int_X L_{\star} (TX)\mid_{\text{vol}},
\end{equation}

\noindent \emph{wherein $b_+$ (resp. $b_-$) is the number of positive (resp. negative) eigenvalues of the noncommutative inner product on $H^n_{\text{dR}} (X,\mathbb R)$, introduced in (\ref  {hirzebruch-star-pairing} ). Moreover, if $X$ is a $4p$-dimensional manifold, then upon the equation of $\tau(X)=\chi(X)~(\text{mod}~2)$, we also find;}
\begin{equation}
\int_X L_{\star p}(TX)=\int_X e_{\star}(TX)~\text{mod}~2.
\end{equation}
~
\par One can work out the index of an elliptic complex for the corresponding complex manifolds.\\

\par \textbf{Corollary 12;} \emph{Assume that $X$ is also a complex manifold with the Dolbeault operator $\overline{\partial}$. Then, the index of the noncommutative version of the Dolbeault operator for the rolled up vector bundles $\wedge^{(0,\text{even})}T^*X$ and $\wedge^{(0,\text{odd})}T^*X$ is given as;}
\begin{equation} \label {corollary16}
\begin{gathered}
\text{Index}(\overline{\partial}_\star)\\
= (-1)^{n}\int_X  \left( \text{ch}_{\star}( \wedge^{\text{even}}TX^-) -\text{ch}_{\star}(\wedge^{\text{odd}} TX^-)  \right)\star \left( \frac{ \text{Td}_{\star} ( TX^{\mathbb C} ) }{e_{\star}(TX_\star)} \right) \mid_{\text{vol}}=\int_X \text{Td}_{\star n}(TX^+).
\end{gathered}
\end{equation}

\noindent \emph{In addition, upon to \textbf{Corollary 4}, this index coincides with the arithmetic genus of $X$ as;}
\begin{equation} \label {corollary16-1}
\text{Index} (\overline{\partial}_\star)=\int_X \text{Td}_{\star n}(TX^+)=\sum_{k}(-1)^k b^{0,k},
\end{equation}

\noindent \emph{wherein $b^{0,k}=\dim_{\mathbb C} H^{0,k}(X, \mathbb C)$, $0 \leq k \leq n$, are the Hodge numbers.} \\

\par The following corollary is in fact, the noncommutative version of the celebrated Hirzebruch-Riemann-Roch theorem.\\

\par \textbf{Corollary 13;} \emph{Assume that $X$ is a complex manifold and $\mathbb C^K \to E \to X$ is a holomorphic vector bundle. Then, the index of $\overline{\partial}^E_\star$ for the rolled up vector bundles $\frak E^{even}=\wedge^{(0,\text{even})}TX^{\mathbb C} \otimes E$ and $\frak E^{odd}=\wedge^{(0,\text{odd})}TX^{\mathbb C} \otimes E$ is given as;}
\begin{equation} \label {c18}
\begin{gathered}
\text{Index}(\overline{\partial}^E_\star)= (-1)^{n} \int_X  \left( \text{ch}_{\star}( \frak E^{even}) -\text{ch}_{\star}(\frak E^{odd})  \right) \left( \frac{ \text{Td}_{\star} ( TX^{\mathbb C} ) }{e_{\star}(TX )} \right) \mid_{\text{vol}}\\
=\int_X \text{ch}_{\star}(E)\text{Td}_{\star}(TX^+) \mid_{\text{vol}}.~~~~~~~~~~~~~~~~~~~~~
\end{gathered}
\end{equation}

\noindent \emph{Moreover, as a topological invariant, the holomorphic Euler character of $E$ is given as;}
\begin{equation} \label {c182}
  \chi(X,E) =\sum_{k=0}^{2n} (-1)^k \dim_{\mathbb C} H^k(X,E) = \text{Index}(\overline{\partial}^E_\star) = \int_X \text{ch}_{\star}(E) \text{Td}_{\star}(TX^+) \mid_{\text{vol}},
\end{equation}

\noindent \emph{where $H^k(X,E)$, $0 \leq k \leq 2n$, are the sheaf cohomology groups of $E\to X$.}\\

\par Suppose that $X$ is a spin manifold and $S(X)=S_(X)+ \oplus S_-(X) \to X$ is its spinor bundle, wherein $S_+(X)$ (resp. $S_-(X)$) is the corresponding subbundle with positive (resp. negative) chirality \cite{poor, nakahara}. Let $D=D_++D_-$ be the Dirac operator for $D_+:\Gamma(S_+(X)) \to \Gamma(S_-(X))$, and $D_-={D_+}^\dag$. Hence, we readily have a noncommutative elliptic complex as;
\begin{equation} \label {dirac com}
0 \xrightarrow[\text{~}]{0}  \Gamma(S_+(X))  \xrightarrow[\text{~}]{D_{+ \star}}  \Gamma(S_-(X))  \xrightarrow[\text{~}]{0} 0.
\end{equation}

\par Moreover, if $\mathbb C^K \to E \to X$ is a complex vector bundle, then the noncommutative Dirac operator is generalized to twisted spinor bundles as $D_{E \star}=D_{+\star} \otimes 1$ and provides the following noncommutative elliptic complex;
\begin{equation} \label {twisted dirac com}
0 \xrightarrow[\text{~}]{0}  \Gamma(S_+(X)\otimes E)  \xrightarrow[\text{~}]{D_{E \star}}  \Gamma(S_-(X)\otimes E)  \xrightarrow[\text{~}]{0} 0,
\end{equation}

\noindent The next corollary has important applications in noncommutative Einstein-Yang-Mills theory, which in fact, extracts the total amount of axial anomaly for a noncommutative $SU(N)$-gauge theory in presence of gravity.\footnote{See \cite{varshovi1, varshovi2, varshovi3} for more discussions on flat spacetime.} \\

\par \textbf{Corollary 14;} \emph{If $X$ is a spin manifold, then the index of $D_+ \star$, the so called noncommutative Dirac index, is given by noncommutative $\hat A_\star(TM)$ genus as \cite{atiyah84};}
\begin{equation} \label {dirac index}
\begin{gathered}
\text{Index}(D_{+\star})= (-1)^{n} \int_X  \left( \text{ch}_{\star}( S_+(X)) -\text{ch}_{\star}(S_-(X) )  \right) \left( \frac{ \text{Td}_{\star} ( TX^{\mathbb C} ) }{e_{\star}(TX )} \right) \mid_{\text{vol}}\\
=\int_X \hat A_{\star}(TM)\mid_{\text{vol}},~~~~~~~~~~~~~~~~~~~~~~~~~~~~~~~~~~~~~
\end{gathered}
\end{equation}

\noindent \emph{which obviously vanishes for odd $n$. Moreover, if $E\to X$ is a complex vector bundle then the index of $D_{E\star}$ is calculated as \cite{getzler, nakahara}:}
\begin{equation} \label {twisted dirac index}
\text{Index}(D_{E\star})=\int_X \text{ch}_\star(E)\hat A_{\star}(TM)\mid_{\text{vol}}.
\end{equation}


\section{Summary and Conclusions}
\setcounter{equation}{0}

\par In this paper we constructed a covariant star product on a semi-conformally flat noncommutative product manifold $X=Y\times Z$, wherein $Y$ is a closed and orientable even dimensional manifold and $Z$ is a flat Calaby-Yau $m$-fold. We generalized the definition of the covariant star product $\star$ to any vector bundle and extracted a well-defined gauge covariant noncommutative curvature $F_\star$ due. Actually, the noncommutativity matrix in $F_\star$ is rescaled by the ordinary curvature elements and this provides a consistent framework for studying noncommutative field theories including that of Yang-Mills and of gravity, within a dynamical noncommutativity matrix, as was pointed out in \cite{steinacker}. 

\par Then, it was shown that by inserting $F_\star$ into symmetric invariant polynomials one could provide many fruitful noncommutative counterparts of characteristic classes. Next, a noncommutative version of Chern-Weil theorem, \textbf{Theorem 1}, was also established for covariant star product $\star$ which states: \textbf{a)} The de Rham cohomology class of noncommutative characteristic classes are independent of the corresponding connection, and; \textbf{b)} The noncommutative characteristic classes are cohomologous to their ordinary (commutative) versions. The correspondence of our formulation and the Seiberg-Witten map is discussed and it was used for providing a second proof for \textbf{Theorem 1}.

\par Afterwards, the noncommutative of elliptic operators was studied and a noncommutative statement of the Atiyah-Singer index theorem on compact manifold \cite{atiyah1}, \textbf{Theorem 2}, was stablished on $X$. This, noncommutative index theorem could be compared with the algebraic Conne's index theorem \cite{connes} and the Fedosov's formula of index theorem on noncommutative symplectic manifolds \cite{fedosov}. The former was studied thoroughly in \cite{varshovi2, varshovi3}, but the latter one could be compared with a generalization of our noncommutative machinery for non-associative vector bundles. However, since the covariant star product will not touch the topological/algebraic structures of vector bundles we readily conclude that a $K$-theoretic viewpoint could be employed to our noncommutative index theorem, via a similar approach used in \cite{ atiyah68-1}.



\section{Acknowledgments}

\par The author says his gratitude to S. Ziaee who was the main reason for appearing this article. Also, it should be noted that this research was in part supported by a grant from IPM  (No.1400810418).




\end{document}